\documentclass{aa}  

\usepackage{graphicx}	% Including figure files
\usepackage{amsmath}	% Advanced maths commands
\usepackage{amssymb}	% Extra maths symbols
\usepackage{txfonts}
\usepackage{lipsum}
\usepackage{lscape}             % to rotate a single page table, example in appendix.
\usepackage{placeins}           % useful with \FloatBarrier, to keep 
\usepackage{xcolor}
\usepackage{hyperref}
\begin{document}

%%%%%%%%%%%%%%%%%%% TITLE PAGE %%%%%%%%%%%%%%%%%%%

% Title of the paper, and the short title which is used in the headers.
% Keep the title short and informative.
    \title{Testing cosmological structure formation in a Unified Dark Matter-Energy model with fast transition}

   \author{Diogo Castel\~ao\inst{1,2}\fnmsep\thanks{\email{dmcastelao@ciencias.ulisboa.pt}}
        \and Alberto Rozas-Fernández\inst{1,3}
        \and Ismael Tereno\inst{1,2}
        }

   \institute{Instituto de Astrof\'isica e Ci\^encias do Espa\c{c}o, Faculdade de Ci\^encias,
Universidade de Lisboa, Tapada da Ajuda, PT-1349-018 Lisboa, Portugal
            \and Departamento de F\'isica, Faculdade de Ci\^encias, Universidade de Lisboa, Edif\'icio C8, Campo Grande, PT1749-016 Lisboa, Portugal
            \and Departamento de Física, Universidad de Castilla-La Mancha, Avenida de Carlos III, s/n, 45004 Toledo, Spain\\}

   \date{Received June 10, 2026}

    \abstract
    {Unified Dark Matter-Energy models (UDM), a class of models where dark matter and dark energy exist as a single cosmological fluid, are an alternative approach to $\Lambda$CDM. In this work we focus on a model with a fast transition between dark matter-like and dark energy-like behaviour. The epoch and rapidity of the transition are the key features to enable the formation of structure in this model. We have studied its viability using CMB and Weak Lensing data with nested sampling inference methods.  We found that the preferred region of the parameter space %with highest probability 
    is the one with early and fast transition models, where it also lies the model's $\Lambda$CDM limit. Our study confirms that this UDM model is able to form cosmological structure compatible with the data used.}
%We find that our UDM model is consistent with structure formation data, and in addition it alleviates the 
%$\Lambda$CDM $\sigma_8$ tension.


    \maketitle

\nolinenumbers

%%%%%%%%%%%%%%%%%%%%%%%%%%%%%%%%%%%%%%%%%%%%%%%%%%

%%%%%%%%%%%%%%%%% BODY OF PAPER %%%%%%%%%%%%%%%%%%

\section{Introduction}

There are several theoretical ideas, alternatives to $\Lambda \mathrm{CDM} $, proposing various forms of dark energy (DE) to explain the origin of the late-time acceleration of the Universe. $\Lambda \mathrm{CDM} $ may be considered as the simplest solution, borrowing Einstein’s idea of vacuum energy, namely the cosmological constant $\Lambda$. However, there are two problems that arise with the cosmological constant and motivate us to study other alternatives. The first is called the fine-tuning problem, and the second is the coincidence problem \citep{DE}.

In Quantum Field Theory (QFT), the notion of empty space has been replaced by a vacuum state, defined to be the ground state of a collection of quantum fields (meaning the lowest energy density). These quantum fields exhibit zero-point fluctuations everywhere in space. These zero-point fluctuations of the quantum fields, as well as other ‘vacuum phenomena’ of QFT, give rise to an enormous vacuum energy density $\rho_{vac}$. This vacuum energy density is believed to act as a contribution to the cosmological constant $\Lambda$. Several observations show that $\Lambda$ is in fact very small, $ \mid \Lambda \mid < 10^{-56} cm^{2}$. This constraint can be interpreted as a constraint on the vacuum energy density in QFT, $\mid \rho_{vac} \mid < 10^{-29} g/cm^{3} \sim 10^{-47} \rm GeV^{4}$. However, one can theoretically estimate the various contributions to the vacuum energy density in QFT, and the predictions exceed the observational bound by at least 40 orders of magnitude. This large discrepancy is the main problem associated with the cosmological constant. The second problem addresses the single question of why $\rho_{\Lambda}$ is not only small but of the same order of magnitude of the matter energy density present in the Universe, $\rho_{m}$.

One of the first alternative models that tried to solve these problems was quintessence  \citep{Wetterich:1987fm, Ratra:1987rm, Caldwell:1997ii, Zlatev:1998tr, Caldwell:2005tm}. The name quintessence means the fifth element, besides baryons, DM, radiation, and spatial curvature. This fifth element is the missing cosmic energy density component with negative pressure that we are seeking today. The basic idea of quintessence is that DE is in the form of a time varying scalar field that is slowly rolling down towards its potential minimum.
The quintessence model assumes both a canonical kinetic energy and a potential energy term in the action. By modifying this canonical kinetic energy term, one can get a non-canonical (non-linear) kinetic energy of the scalar field that can drive the negative pressure without the need for potential terms. The non-linear kinetic energy terms are thought to be small and are usually ignored because the Hubble expansion damps the kinetic energy density over time. However, it is possible to have a dynamical attractor solution which forces the non-linear terms to remain non-negligible. This kind of models is called k-essence \citep{Chiba:1999ka, Armendariz-Picon:2000nqq, Armendariz-Picon:2000ulo}.

There are other types of models such as the Coupled Dark Energy models that consider an interaction between DM and DE \citep{Wetterich:1994bg, Amendola:1999er}, or f(R) gravity \citep{Capozziello:2003tk, Carroll:2003wy, Nojiri:2010wj} that resort to the modification of gravity on large scales, or even other ideas like the DGP model \citep{Dvali:2000hr, Deffayet:2000uy} and the inhomogeneous LTB model \citep{Tomita:2000jj, Alnes:2005rw}. See \citep{2025PDU....4901965D} for a recent review of this plethora of models.

One of the most elegant alternatives, which is the one we consider in this work, is that of the Unified Dark Matter-Energy models (UDM), also known as quartessence. In these models, a single fluid component behaves as dark matter (DM) at high redshifts and as DE at low redshifts. This approach also has the advantage of evading the coincidence problem. Most of these models are characterised by a sound speed that for non-zero values imprints oscillations on the matter power spectrum, inhibiting structure growth at late times on those scales that are smaller than its corresponding Jeans length.

The first UDM model proposed was the Generalised Chaplygin Gas (GCG) \citep{Kamenshchik:2001cp, Bento:2002ps}. However, this model has a speed of sound which may become significantly large during the evolution of the Universe preventing the formation of structure \citep{Sandvik:2002jz}. In order to make the GCG viable, four main solutions have been proposed: the vanishing sound speed models, also known as silent Chaplygin Gas \citep{Amendola:2005rk}; the decomposition models \citep{Wands:2012vg}, where the UDM models are seen as an interaction between DM and the vacuum energy and it is possible to build a model with zero speed of sound; the clustering GCG model \citep{Kumar:2014pia}; and also a model where considering the backreaction of the small scale structures on the large scale evolution
of the Universe has the effect of making the model consistent with observations provided that the non-linear clustering level is high enough \citep{Avelino:2014nva}.

Besides the different possibilities considered for the CGC many other UDM models were proposed \citep[see e.g.][]{Scherrer:2004au}. Some of them have a non-canonical kinetic term in the Lagrangian (a kinetic term $f(\dot{\varphi}^{2})$ instead of $\dot{\varphi}^{2}/2 $) that allows one to build a model with a small effective sound speed and eventually form structure. Another  alternative are the UDM models with a fast transition. These models have a fast transition between a CDM-like epoch, with an Einstein-de Sitter evolution, and an accelerated DE-like epoch. This fast transition results in having a speed of sound different from zero but only for a short period of time and therefore a Jeans scale large enough for structure to form.

By prescribing an evolution for the equation of state, $w$, for the pressure $p$, or for the energy density $\rho$, we can define the dynamics of the UDM model. The first UDM model with a fast transition was proposed in \citep{Piattella:2009kt}, where the evolution of $p$ was prescribed. In this model, the UDM fluid was considered to be barotropic, $p=p(\rho)$, and the perturbations were adiabatic.\ A second UDM model with a fast transition was presented in \citep{Bertacca:2010mt}. This model was built from a k-essence scalar field. In this model it was also prescribed an evolution for $p$ but non-adiabatic perturbations were considered. Here we consider a third UDM model first proposed in \citep{Bruni:2012sn}. The dynamics of this UDM fluid is prescribed through the energy density $\rho$, and it has adiabatic perturbations.\ This model was presented as a phenomenological model, with no discussion about the physical process behind the fast transition.

In order to fully describe the properties of our UDM model of interest, we start in section 2 with the detailed description of the model. In section 3 we move to the model testing, performing several Bayesian inference analyses using Markov Chain Monte Carlo (MCMC) Nested Sampling methods with various sets of cosmological structure formation and background expansion data. We conclude in section 4, showing for the first time the viability of a UDM model with a fast transition at structure formation level, implying that the UDM approach is a possible candidate to describe DM and DE.

\section{Model}

\subsection{Model description}

We assume a spatially flat Friedmann-Lemaitre-Robertson-Walker (FLRW) background geometry, with metric:
\begin{equation}
ds^{2} = -dt^{2}+a^{2} \left[dr^2+r^{2}(d \theta^{2}+\sin(\theta )^{2} d \varphi^{2}) \right],
\end{equation}
where $a(t)$ is the scale factor and $t$ is the cosmic time. In addition to the cosmic time $ t $, it is also convenient to introduce the conformal time defined by:
\begin{equation}
\tau \equiv \int a^{-1} dt\,.
\end{equation}

The total energy-momentum tensor is that of a perfect fluid with energy density $\rho$, pressure $p$ and four-velocity $u_\mu$: $T_{\mu \nu} = (\rho + p) u_\mu u_\nu + p g_{\mu \nu}$. Starting from these assumptions, and choosing units such that $8\pi G = c = 1$, the Einstein equations take the usual form:
\begin{equation}
\left(\dfrac{\dot{a}}{a} \right)^{2}=\dfrac{\rho}{3},
\end{equation}
\begin{equation}
\dfrac{\ddot{a}}{a}=-\dfrac{1}{6}(\rho + 3p)\,
\end{equation}	
and the conservation equation is given by the continuity equation:
\begin{equation}
\dfrac{d}{dt}(a^{3} \rho)+ p\dfrac{d}{dt}(a^{3})=0\,.
\end{equation}

The components of the cosmological fluid considered in this scenario are radiation, baryonic matter and the UDM fluid that accounts for both DM and DE. The energy density of the UDM fluid is modelled such as to have a $a^{-3}$ dark matter-like evolution at early times. At a transition scale factor, $a_t$, a second term is switched on, slowing down the decreasing rate of the density that eventually becomes dominated by a DE behaviour and reaches a constant value $\rho_{\Lambda udm}$. In particular, our prescription is \citep{Bruni:2012sn}:
\begin{equation}
\rho= \rho_{t} \left( \dfrac{a_{t}}{a} \right)^{3} + \rho_{\Lambda {\rm udm}} \left[ 1 - \left( \dfrac{a_{t}}{a} \right)^{3} \right] H(a-a_{t})\,,
\label{rhoudm}
\end{equation}
where H(a) is the Heaviside function. The UDM density is thus parameterised by $a_{t}$, the scale factor at the transition, and the two density parameters $\rho_{t}$, the value of $\rho$ at $a=a_{t}$ and $\rho_{\Lambda udm}$, the limiting value of $\rho$ at infinity. The number of independent parameters is reduced to two due to the constraint  imposed by the Friedmann equation. For example, $\rho_{t}$ can be determined from the other parameters: 
\begin{equation}
\rho_{t} = \frac{\rho_c - \sum_{i} \rho_{0,i} - \left[\rho_{\Lambda udm} - \rho_{\Lambda udm}a_{t}^{3} \right]}
{a^3_{t}}\,,
\end{equation}
where $\sum_{i} \rho_{0,i}$ is the contribution from the various components of the Universe (baryons, photons, or any other non-UDM component) at present time, and $\rho_c$ is the present critical density. For convenience and for a better comparison with the $\Lambda$CDM scenario, we may refer to the first term on the right-hand side of Eq.~(\ref{rhoudm}) as a matter density $\rho_{\rm m}$ and to the second term as a DE density. This DE density today is as close to $\rho_{\Lambda {\rm udm}}$ as $\rho_t$ is larger, {\em i.e.} as the transition happens earlier. 

Modelling the UDM fluid through its density allows us to write the equation of state and the adiabatic sound speed as functions of the density and its derivatives only. Using the continuity equation we obtain:
\begin{equation}
w = - \dfrac{a}{3} \dfrac{\rho^{'}}{\rho} -1
\label{eos}
\end{equation}
and
\begin{equation}
c_{s}^{2} = \frac{dp}{d\rho} = \dfrac{a}{3} \dfrac{\rho^{''}}{\rho^{'}} - \dfrac{4}{3}\,,
\label{soundspeed}
\end{equation}
where the derivatives are taken with respect to conformal time. The model is thus completely specified by the density.

For the energy density to be a continuously differentiable function, we need to specify an approximation to the Heaviside function. There are many known approximations to the Heaviside function \citep{Bracewell:2003}, and they usually introduce an extra parameter to the model: the rapidity of the transition, $\beta$. A functioning UDM model with a fast transition has thus two additional parameters with respect to $\Lambda$CDM: $a_t$ and $\beta$.
Choosing different approximations to the Heaviside function may result in different behaviours, since the second-order derivatives of $\rho$ may be quite different in the various cases, producing different evolutions of the sound speed and consequently affecting structure formation in different ways. In all cases the density has a smooth evolution, even for fast transitions (meaning high values of $\beta$), while the evolution of the equation of state and of the sound speed show a sharper transition. As a general behaviour, the speed of sound shows a sharp increase of amplitude around the time of transition in the form of a peak.
The rapidity of the transition impacts both the amplitude and the width of the peak, with faster transitions producing higher and narrower peaks. 

After the peak, the evolution of the sound speed depends on the approximation to the Heaviside function used. For some functions, such as the {\em logistic} function, the sound decreases to negative values before returning to zero. The impact on structure formation is an exponential growth of the density contrast, ruled out by observations \citep{Beca:2005gc}. 

In this paper, we use the {\em atanoriginal} approximation \citep{Bruni:2012sn},
\begin{equation}
H(a - a_{t}) = \dfrac{1}{2} + \dfrac{1}{ \pi } {\rm arctan}(\beta (a - a_{t}) )\,.
\label{hatanoriginal}
\end{equation}
\begin{figure}%[h]
	\centering
	\includegraphics[width=1.0\hsize]{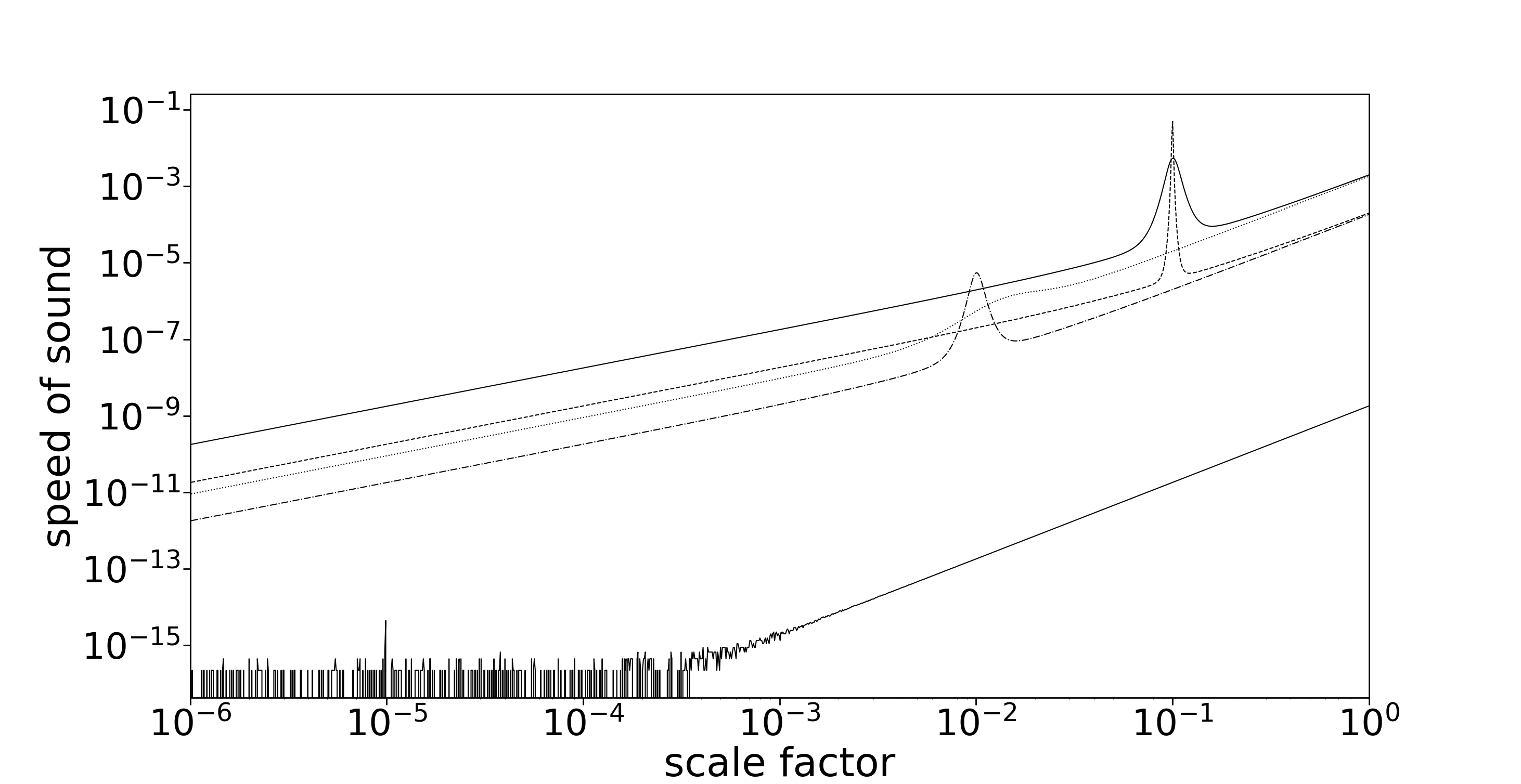}
	\caption
	{Evolution of the speed of sound in the {\em atanoriginal} model, for the following values of the $(a_t\,,\beta)$ pair: $(0.01,10^2)$ (dotted line), $(0.01,10^3)$ (dash-dotted line), $(0.1,10^2)$ (solid line), $(0.1,10^3)$ (dashed line), $(10^{-5},10^8)$ (solid line in the bottom, this case is close to the $\Lambda$CDM limit). The third UDM parameter, $\rho_{\Lambda{\rm udm}}$, is fixed at $0.683\,\rho_{\rm c}$.}
\label{atanoriginal}
\end{figure} 
With this prescription the sound speed continues to increase after the peak. The behavior is shown in Fig.~\ref{atanoriginal} for different combinations of time and rapidity of transition. For early transitions, the increase in the sound speed at the transition is small, while for later transitions the peak can reach large amplitudes especially for fast transitions.
For any time of the transition, the overall increase is larger in the case of slow transitions, where a broad peak takes the amplitude to a higher value than in the case of fast transitions that reach temporarily higher amplitudes through a sharper peak, but then return to lower amplitudes. 
As we can already infer, the viability of these models will be determined by a trade-off between the two parameters $a_t$ and $\beta$. Lower amplitudes of the speed of sound are obtained in models with early and fast transitions, tending to $\Lambda$CDM when $a_t \to 0$ and $\beta \to \infty$. This UDM prescription has the potential to provide a broad range of valid models, distinguishable from $\Lambda$CDM, while having a $\Lambda$CDM limit. 

\begin{figure*}
\centering{\hbox{%
    \includegraphics[width=0.5\hsize]{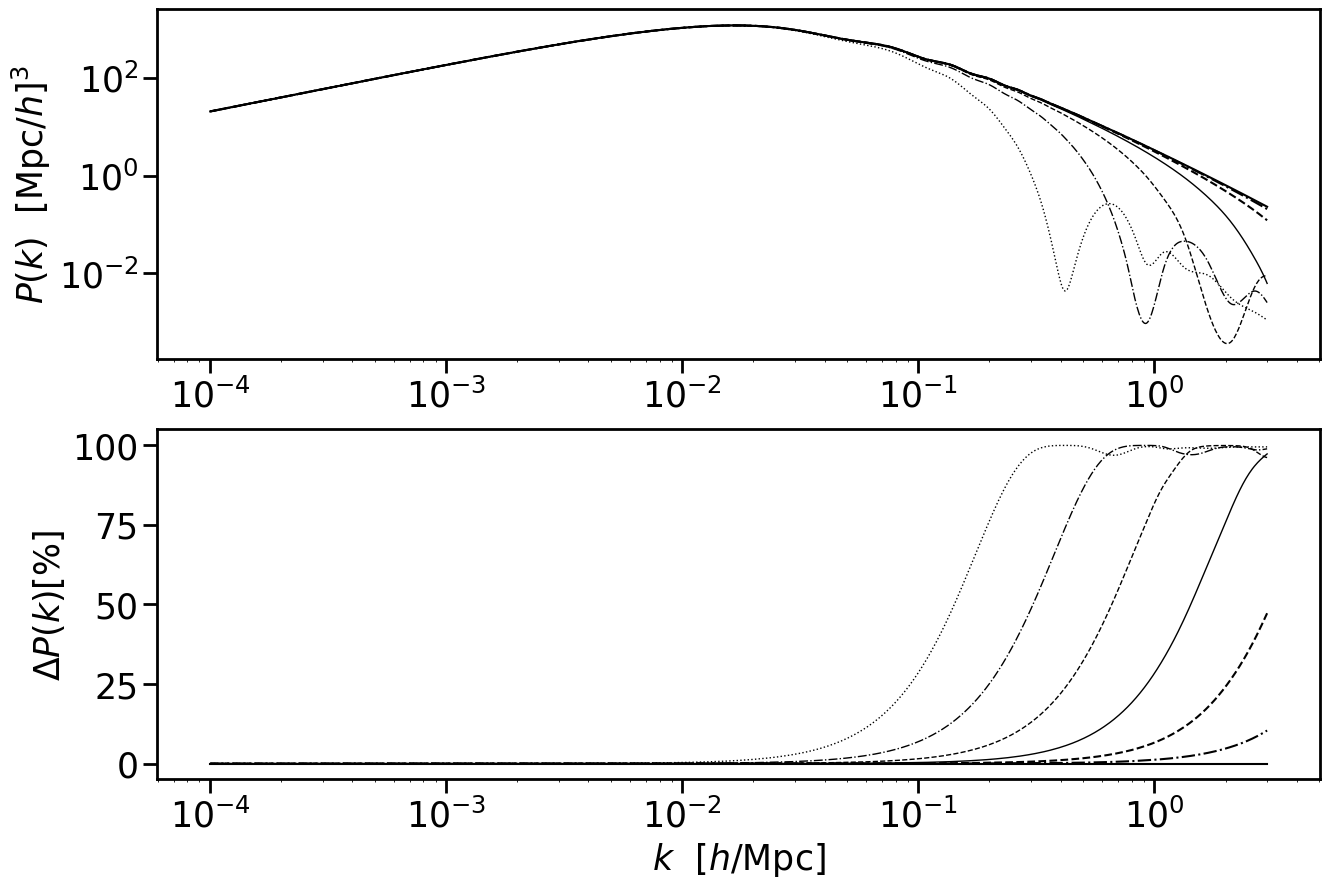}
    \includegraphics[angle=0,width=0.5\hsize]{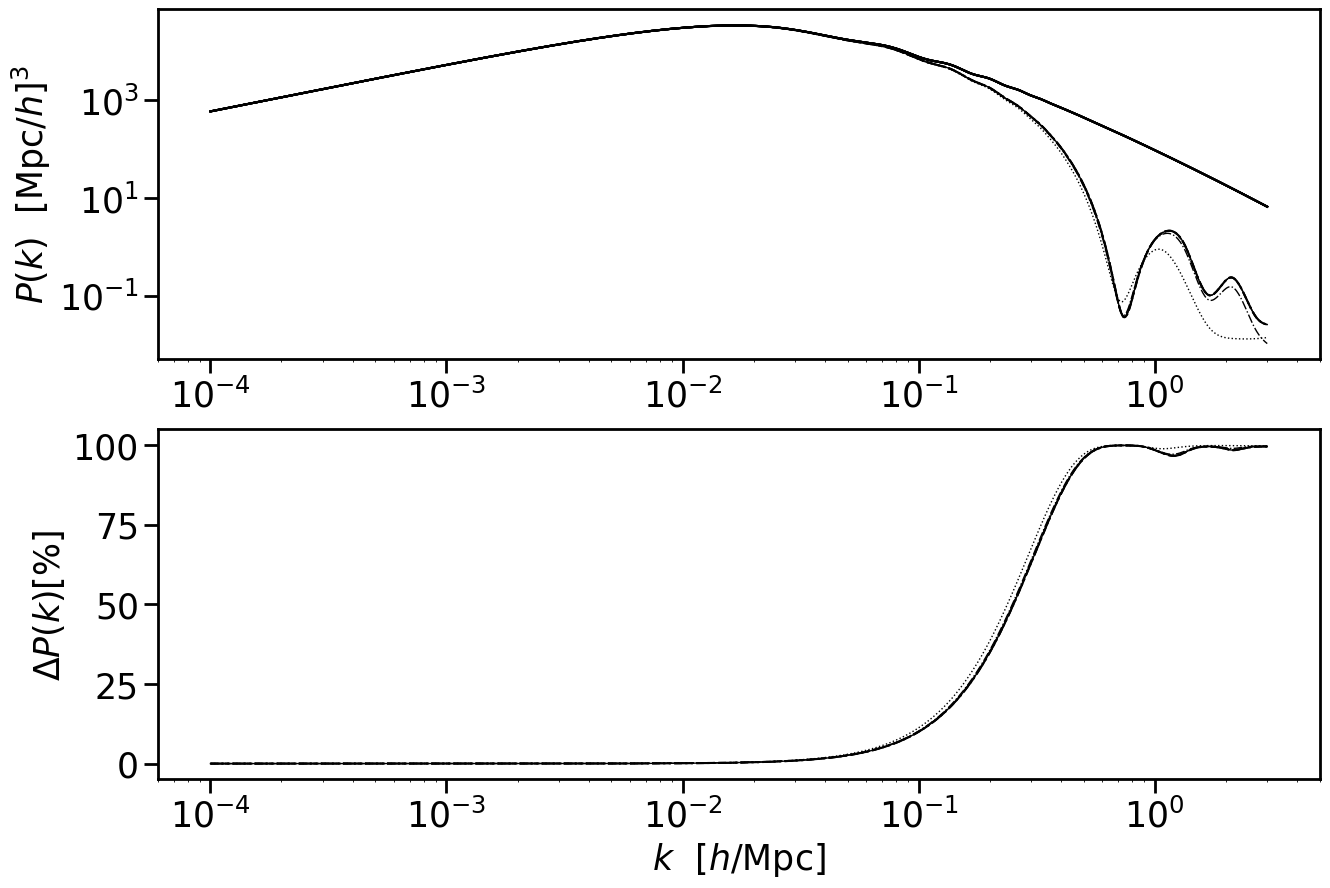}}}
	\caption{UDM matter power spectrum and its relative deviation to the reference $\Lambda$CDM (see text). {\em Left panel}: Dependence on $\beta$; $a_t$ is fixed at $10^{-3}$, $\beta$ varies from $10^4$ to $10^8$ in six equally-spaced logarithmic steps. $\beta=10^8$ (solid line) is indistinguishable from the reference $\Lambda$CDM model. The deviation from the reference model decreases monotonically with $\beta$. {\em Right panel}: Dependence on $a_t$; $\beta$ is fixed at $10^{4.5}$, $a_t$ varies from $10^{-4.5}$ to $10^{-1.5}$ in three equally-spaced logarithmic steps. $a_t=10^{-4.5}$ (solid line) is indistinguishable from the reference $\Lambda$CDM model. The deviation from the reference model increases monotonically with $a_t$.}
\label{powerspectra_matter}
\end{figure*}

\subsection{Linear perturbations}

We are interested in studying structure formation in the UDM scenario. For that, we consider the evolution of scalar perturbations in the conformal Newtonian gauge, where the metric writes:
\begin{equation}
ds^{2}=a^{2}(\tau)\left[- \left(1+2\Psi \right) d\tau^{2} + (1-2\Phi)\delta_{ij}dx^{i}dx^{j} \right]\,.
\end{equation}

We obtain the evolution of the perturbations in the UDM fluid through the following set of perturbed Einstein and conservation equations:

\begin{gather}
3\mathcal{H} \left(\mathcal{H}\Psi - \Phi' \right) + k^{2} \Phi = -4 \pi G a^{2} \rho \delta\,,
\label{einstein00}
\\
k^{2} \left(\Phi' - \mathcal{H}\Psi \right) = 
%4 \pi G a^{2} (\rho + P) \theta = 
4 \pi G a^{2} (1 + w)\rho \theta\,,
\label{einstein0i}
\\
\Psi = - \Phi\,,
\label{einsteinij}
\\
\Phi'' + 2 \mathcal{H} \Phi' - \mathcal{H} \Psi' - \left( \mathcal{H}^{2} + 2 \mathcal{H}' \right) \Psi  = -4 \pi G a^{2} c_{s}^{2}  \rho \delta \,.
\label{einsteinii}
\end{gather}
\begin{gather}
\delta' + 3 \mathcal{H} \left(c_{s}^{2} - w \right) \delta = - \left( 1 + w \right) \left(\theta + 3 \Phi' \right)\,,
\label{cons1}
\\
\theta' + \left[ \mathcal{H} (1 - 3w) + \frac{w'}{1+w} \right] \theta = - k^{2} \left( \dfrac{c_{s}^{2}}{1 + w} \delta	+ \Psi \right)\,,
\label{cons2}
\end{gather}
 where $w$ and $c_s$ are respectively the equation of state and the sound speed of the UDM fluid, $\delta$ is the UDM fluid density contrast, $\theta$ is the velocity perturbation, $\mathcal{H}$ is the conformal Hubble function, $k$ is the Fourier scale and a prime denotes derivative with respect to conformal time.

We can gain insight on the properties of the model by considering the equation of motion for $\Phi$, which is obtained by combining Eq.~(\ref{einstein00}), (\ref{einsteinij}) and (\ref{einsteinii}), yielding:
\begin{equation}
\Phi'' + 3\mathcal{H} \left( 1 + c_{s}^{2} \right) \Phi' +
(c_{s}^{2} k^{2} + 3 c_s^2\mathcal{H}^{2} + 2\mathcal{H}'  + \mathcal{H}^{2} ) \Phi = 0\,.
\label{phimotion}
\end{equation}
Introducing in Eq.~(\ref{phimotion}) the variable:
\begin{equation}
u \equiv \dfrac{2 \Phi}{\sqrt{\rho(1+w)}}\,,
\end{equation}
we can eliminate the friction term and obtain \citep{Mukhanov:2005sc}:
\begin{equation}
u'' + \left( k^{2} c_{s}^{2} - \frac{\alpha''}{\alpha} \right) u = 0\,,
\label{umotion}
\end{equation}
where
\begin{equation}
\alpha \equiv \sqrt{\dfrac{1}{3(1+w)}} \,\frac{1}{a}\,.
\end{equation}
Eq.~(\ref{umotion}) defines a threshold scale, the Jeans scale \citep{Bertacca:2010mt}, in the form:
\begin{equation}
k_{\rm J}^{2} \equiv \left\vert \frac{\alpha''}{c_{s}^{2} \alpha} \right\vert \,.
\label{jeans}
\end{equation}
The density contrast will grow on scales larger than the Jeans scale, $k < k_{\rm J}$, and will oscillate on smaller scales. Models with large $k_{\rm J}$ are then the most favourable to produce structure. Models with a low sound speed (CDM-like) are obviously plausive for structure formation and, as we may  have guessed, they have a large value of $k_{\rm J}$. 

The Jeans scale of this model, Eq.~(\ref{jeans}), may be computed as a function of the derivatives of $\rho$, revealing explicitly all the dependencies on $w$ and $c_{s}^{2}$. After some calculations we obtain an explicit form for $k_{\rm J}^{2}$, \citep[see also][]{Piattella:2009kt}
\begin{gather}
 k_{\rm J}^{2} =\left|\frac{1}{2}(c_{\rm s}^2 - w) - \rho\frac{dc_{\rm s}^2}{d\rho} + \frac{3(c_{\rm s}^2 - w)^2 - 2(c_{\rm s}^2 - w)}{6(1 + w)} + \frac{1}{3}\right| \times
\nonumber
\\
\times \frac{3}{2} \rho a^2 \frac{(1 + w)}{c_{\rm s}^2} \,.
\label{jeans2}
\end{gather}
From Eq.~(\ref{jeans2}), we conclude that a large Jeans scale (favouring structure formation) is obtained not only when the speed of sound is small but also when the speed of sound changes rapidly. This is the motivation to build UDM models with a fast transition.

\begin{figure*}
\centering{\hbox{%
    \includegraphics[width=0.5\hsize]{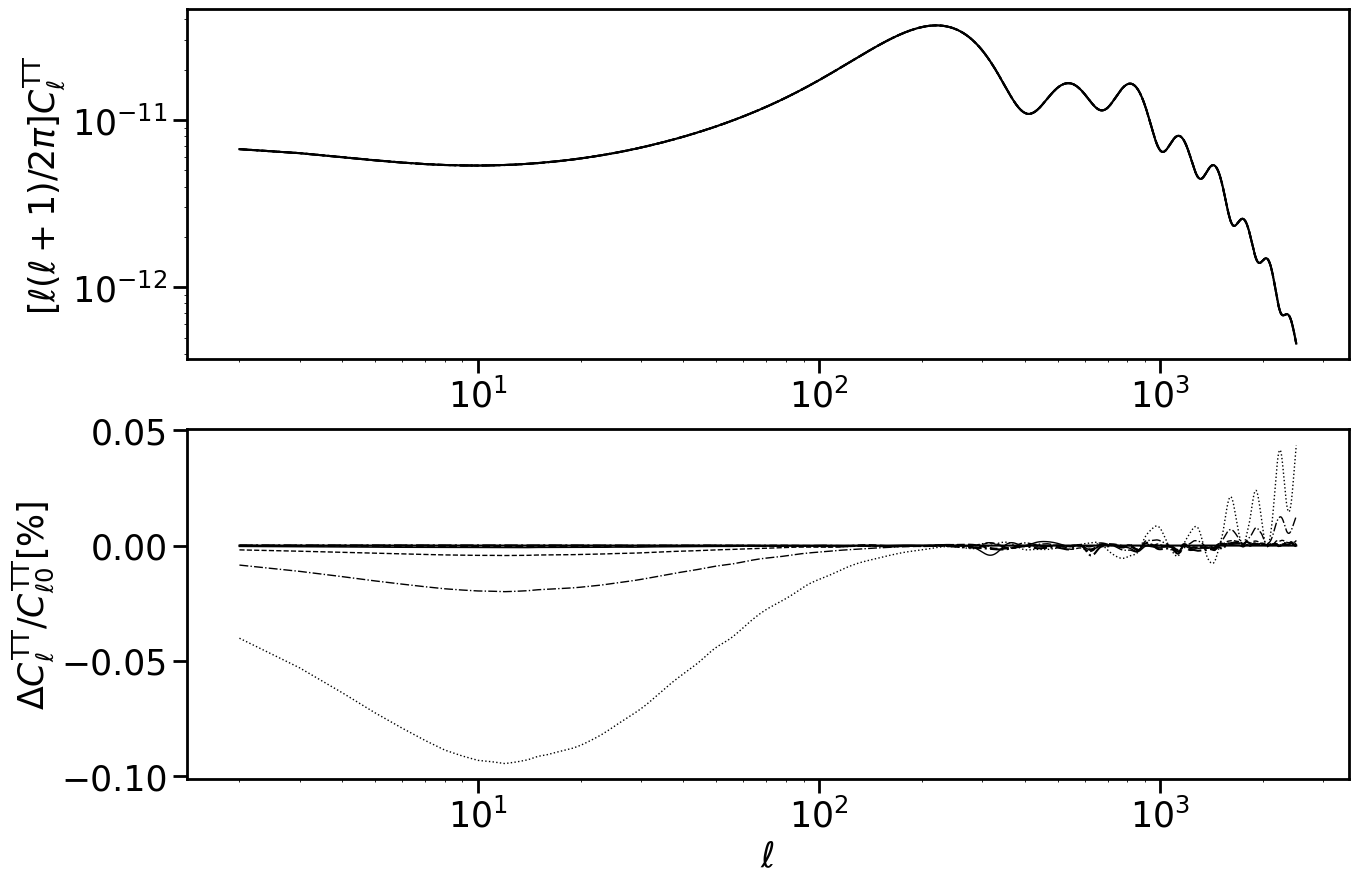}
    \includegraphics[angle=0,width=0.5\hsize]{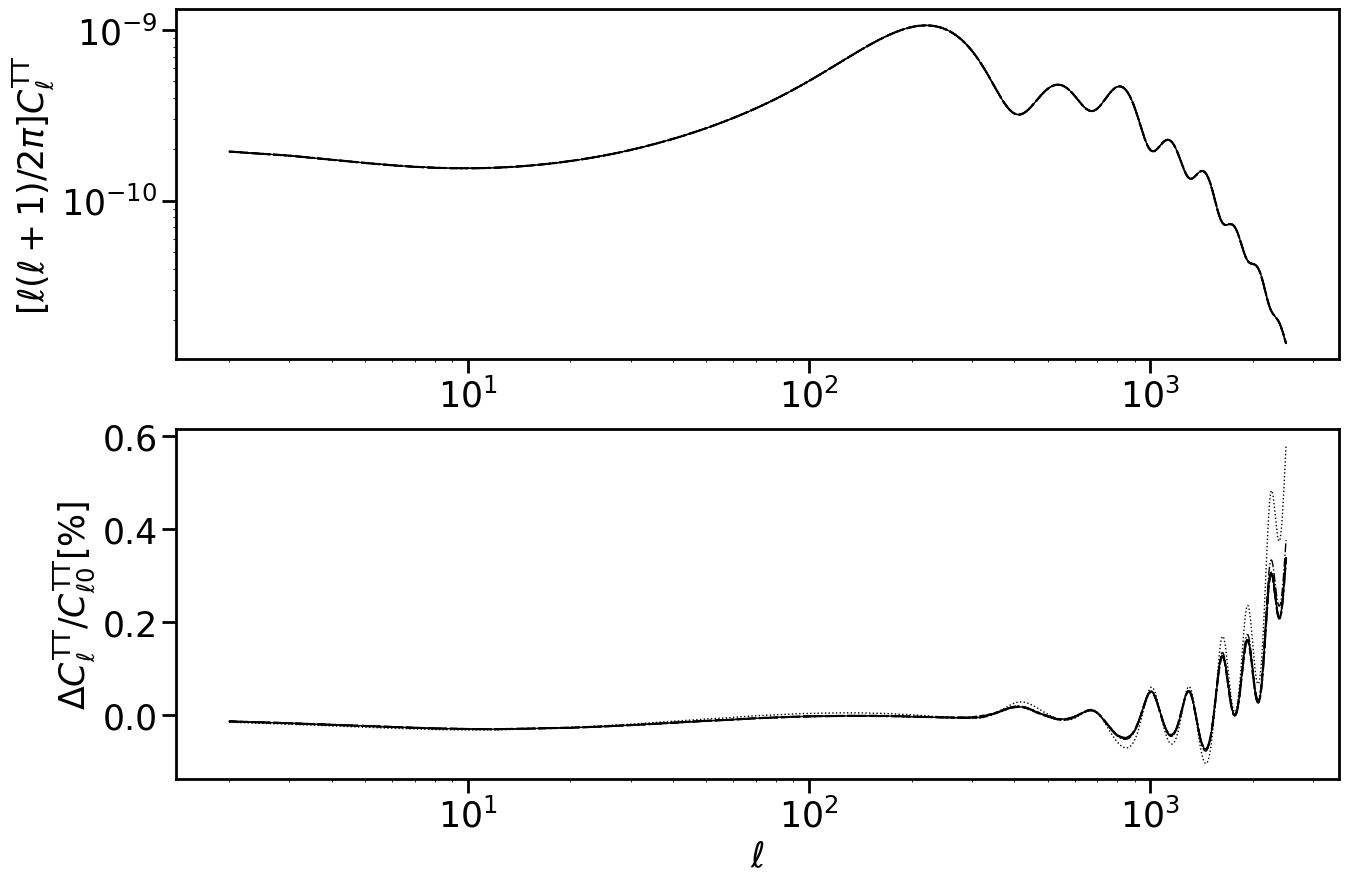}}}
    \caption{UDM CMB TT power spectrum and its relative deviation to the reference $\Lambda$CDM. {\em Left panel}: Dependence on $\beta$ with fixed $a_t$ (same parameter values as in Fig.~\ref{powerspectra_matter}). {\em Right panel}: Dependence on $a_t$ with fixed $\beta$ (same parameter values as in Fig.~\ref{powerspectra_matter}.)}
\label{powerspectra_TT}
\end{figure*}

We implemented our fast transition UDM model in CLASS \citep{Lesgourgues:2011re}. For this we added the contribution of the UDM fluid, defined by the prescriptions given in Eqs.~(\ref{rhoudm}) and (\ref{hatanoriginal}), to the contributions of the radiation and baryonic matter components to the background evolution. To compute the linear evolution of the UDM fluid, we implemented Eqs.~(\ref{cons1}) and (\ref{cons2}) in the synchronous gauge used in CLASS \citep{Ma:1995ey}. The equation-of-state and sound speed, Eqs.~(\ref{eos}) and (\ref{soundspeed}) were derived analytically and the resulting expressions implemented in CLASS. This allowed us to compute the matter and cosmic microwave background (CMB) power spectra. We note that in unified models the matter density contrast is usually defined as the UDM density contrast with respect to the mean matter density \citep{Piattella:2009kt}, {\em i.e.} 
\begin{equation}
\delta_{\rm m}=\delta\,\frac{\bar\rho}{\bar\rho_{\rm m}}\,,
\end{equation}
where $\delta$ is the density contrast of Eq.~(\ref{cons1}), $\bar\rho$ is the background density of the UDM fluid given by Eq.~(\ref{rhoudm}) and $\bar\rho_{\rm m}$ is the matter density given by the first term of Eq.~(\ref{rhoudm}). 
This definition of the matter density contrast provides a consistent matter power spectrum amplitude in the $\Lambda$CDM limit.

Figure~\ref{powerspectra_matter} illustrates the dependence of the matter power spectrum on the model parameters, $\beta$ and $a_t$, for a range of models. 
The third UDM parameter, $\rho_{\Lambda{\rm udm}}$, is kept fixed at $0.683\,\rho_{\rm c}$. The power spectra are compared with a reference flat $\Lambda$CDM model $(\Omega_\Lambda = 0.683, \Omega_{\rm m} = 0.317)$. The models follow closely the reference model on large scales and deviate from it, oscillating, on smaller scales. As the rapidity of the transition decreases, the deviation arises on increasingly larger scales. Likewise, as the transition moves to later times the deviation arises on increasingly larger scales. For the range of values shown, the power spectrum has a similar dependence on both parameters, with deviations from the reference model up to $100\%$ on the smallest scales. 

The CMB TT power spectrum is much less sensitive to the UDM model parameters than the matter power spectrum, as shown in Fig.~\ref{powerspectra_TT}. Besides showing small deviations on small-scales, there is also a sub-percent difference of amplitude on large scales as compared to the reference $\Lambda$CDM model. The effect is more noticeable for slow transition models, where the equation of state of the UDM fluid takes longer to pass from dark matter-like to dark energy-like. Hence, the DE behaviour kicks in later, producing a slower decrease of the amplitude of the gravitational potentials and thus a smaller increase in the energy of the CMB photons, through the integrated Sachs-Wolfe (ISW) effect, as compared with the reference $\Lambda$CDM model.

We note that to confirm there are no numerical artifacts contributing to the oscillations in the matter power spectrum and that they are indeed caused by the variation of the sound speed,  as a test we fixed the sound speed to zero in the perturbation module of \texttt{CLASS}, obtaining a matter power spectrum with no oscillations. 
%This gives rise to possible extensions that could make the model well behaved in a larger region of parameter space.

\section{Testing UDM}

We turn now to the process of constraining the model parameters by comparing its power spectra predictions with data.

\subsection{Datasets}

In our analyses the main source of information on the properties of structure formation in the late Universe is provided by the weak lensing (WL) KiDS+VIKING-450 dataset. We use the publicly available shear 2-pt correlation functions $\xi_\pm$ provided in nine angular bins in the range $l=76$ to $l=1310$ and corresponding to cross-correlations between pairs of five redshift bins from $z \sim 0.3$ to $z \sim 1.0$, as well as the likelihood code from \citep{Hildebrandt:2018yau}. The estimated shear correlation function depends not only on the cosmological parameters but also on a set of nine nuisance parameters that model uncertainties in the baryon feedback, in intrinsic alignments, additive and multiplicative shear measurement bias, and uncertainties in the mean redshifts of the five tomographic bins. We divide the dataset in two parts: WL 'full' containing all scales of the two-point vectors, and WL 'linear' where we cut-out the non-linear scales.

We also use the CMB Planck 2018 data \citep{Planck:2018vyg}. We downloaded the likelihood codes provided by the Planck Team ~\footnote{http://pla.esac.esa.int/pla} 
and installed them in \texttt{MontePython}. 
We use the likelihood \textit{Planck\_highl\_TTTEEE\_lite} on small scales. This lighter version of the likelihood fixes all of the nuisance parameters but one, $A_{planck}$, a free parameter calibrating the overall amplitude of the CMB power spectra. 
On large scales, we use the likelihoods \textit{Planck\_lowl\_EE} and \textit{Planck\_lowl\_TT}. On these scales, the Planck power spectrum is derived from Planck data combined with the nine-year WMAP sky maps \citep{WMAP:2012fli}, and the 408-MHz survey \citep{Haslam:1982zz}, including $93 \%$ of the sky.

\subsection{Setting up the analysis}

The analysis is made with MontePython \citep{Brinckmann:2018cvx}. This software is prepared to work integrated with CLASS and contains already several likelihood codes \citep{Audren:2012wb}.
It provides a variety of sampling methods: MCMC Metropolis-Hastings \citep{Lewis:2002ah}, Nested Sampling \citep{Feroz:2007kg,Feroz:2008xx} MultiNest \citep{Feroz:2013hea} and PolyChord \citep{Handley:2015fda} implementations, EMCEE \citep{Foreman-Mackey:2012any}, CosmoHammer implementation \citep{Akeret:2012ky} and Importance Sampling.\\

The parameter space consists on the three UDM parameters: $a_t$, $\beta$ and $\Omega_{\Lambda {\rm UDM}}$; four cosmological parameters: the amplitude and slope of the primordial power spectrum, $A_{\rm s}$ and $n_{\rm s}$, respectively, the baryon density ($\Omega_{\rm b}$) and the Hubble constant ($h$); as well as the probe-dependent nuisance parameters. The nuisance parameters used were the rescaling $A_{\rm Planck}$ for CMB, the amplitude of the intrinsic alignment model $A_{\rm IA}$ and the baryonic feedback parameter $A_{\rm bary}$ (only for the 'full' analysis) for WL. The remaining cosmological and nuisance parameters were kept fixed, and we only considered flat models. We note that we do not explicitly include a dark matter density parameter but use the $\Omega_{\Lambda \rm UDM}$ parameter instead, which is more convenient in the UDM model. 

We start by making preliminary analyses to establish an optimal setup to be used in producing the final analyses and results.
The analyses were done with flat priors on the UDM parameters. The prior on $\beta$ has an impact on the results, since reducing the upper limit of  $\beta$ excludes the $\Lambda$CDM limit. We tested this limit by also considering a stringent prior on $\beta$ of $[0.01,10^{5.7}]$. Since one feature found in the results is a strong $(\beta, \sigma_8)$ degeneracy, the use of a stringent prior artificially lowers the mean of the $\sigma_8$ distribution. We studied the variation of the $\sigma_8$ constraint with the $\beta$ prior and found that an upper limit of $10^{12}$ is required to obtain a robust $\sigma_8$ estimate. The final analyses were performed using the UDM parameter flat prior ranges shown in Table~\ref{tab: udm_priors}.

\begin{table}
\centering
\caption{Ranges for the UDM parameters used in the nested sampling analyses.} 
%\vspace*{0.5cm}
\begin{tabular}{|l|c|c|c|} 
 \hline 
  & $a_{t}$ & $\beta$ & $\Omega_{\Lambda \rm UDM}$ \\ \hline 
prior &$10^{-5}$ - $10^{0}$ & $0.01$ - $10^{12}$ & $0.01$ - $1$ \\ 
\hline 
 \end{tabular} \\
\label{tab: udm_priors}
\end{table}

To be able to test the UDM model we had to proceed carefully by implementing some tailor-made modifications to the analysis tools. In the large $\beta$ regime, the peak of the speed of sound becomes very narrow;  the resolution of the scale factor timestep needs to be increased to ensure the peak is accounted for and a correct matter power spectrum is computed by \texttt{CLASS}. 

\begin{figure}
	\centering
\includegraphics[width=0.95\hsize]{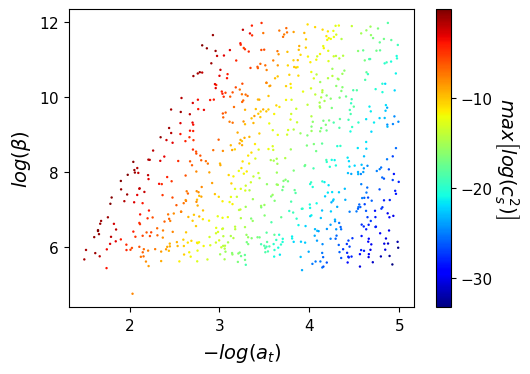}
	\caption {Three-dimensional plot showing the maximum speed of sound for a large sample of models in the $\beta$ and $a_t$ plane. The rapidity of the transition is shown in the y-axis increasing from bottom to top; the time of transition is shown in the x-axis increasing from right to left; the colors indicate the maximum speed of sound increasing from blue to red}.
\label{UDM_cs2_max}
\end{figure}

We also had to deal with models for which the peak reaches large values, $c_s^2 > 1$, for which \texttt{CLASS} fails to produce a result. In our analyses, we restricted the speed of sound to values lower than one. 
Figure~\ref{UDM_cs2_max} shows how the peak of the sound speed varies with $\beta$ and $a_t$. The peak amplitude depends on a combination of the two parameters; the plot showing degeneracy bands from the upper right side (fast and early transitions) to the lower left side (slow and late transitions). The highest amplitudes would lie in the upper left corner, that remains empty since the corresponding models with fast and late transitions are the ones with $c_s^2 > 1$ discarded in our setup. Of the remaining models, the ones that produce larger, unwanted, oscillations in the matter power spectrum are the slow and late transitions in the lower left corner. 
We verified that the oscillations are indeed real and not a numerical artifact by comparing the results obtained with different numerical integrators. 
In summary, the speed of sound is higher for fast and late transition models and lower for early and slow transition ones, in agreement with the examples seen in Fig.~\ref{atanoriginal}. This is not in contradiction with the theoretical argument, following from Eq.~\eqref{jeans2}, that a fast transition favours structure formation. Indeed, if the maximum value of the speed of sound is sufficiently low, a fast transition produces less oscillations than a slower one, even with a potentially higher maximum value, as we will see in our analysis.

We realized that given the sparse sampling of scales provided by the KiDS data, it happens that some models with large oscillations provide spurious good fits to the sparse data. This produces a multi-peaked likelihood with high likelihood regions scattered on the parameter space. The standard Metropolis-Hastings MCMC algorithm proved unable to sample this likelihood function, since it easily got stuck in local peaks. To overcome this problem we used PolyChord \citep{Handley:2015fda}, a nested sampling algorithm. In contrast to MCMC, nested sampling methods do not sample the multi-dimensional likelihood in one sequence over the full range of likelihood values, but in a series of sequences within likelihood thresholds.    
This approach is better suited to deal with multi-peak distributions. Moreover, the fact that nested sampling methods are capable of sampling a broad likelihood range at similar resolutions for all likelihood bins, allows  it to compute the integral of the full distribution with enough resolution on the tails, which in contrast would be poorly sampled with MCMC. This provides a measure of the evidence, which we will use for model comparison.

Another point to consider when setting up the analysis is whether the scale range of weak lensing data falls in the linear or non-linear regime. We used $\Lambda$CDM as a benchmark and computed its shear two-point correlation functions $\xi_\pm$ from linear and non-linear theoretical matter power spectra, with the non-linear correction computed using the HMCODE~\citep{Mead:2016zqy}. Comparing the linear and non-linear predictions across the various redshift bin cross-correlations, we identified the scales at which they deviate, thereby defining a threshold scale for each two-point function. This allowed us to flag non-linear points in the KiDS dataset.

We performed separate ‘linear’ and ‘full’ analyses. In the 'linear' case, we removed the points identified as non-linear. In the 'full' case, we kept all data points and applied the HMCODE method to estimate the non-linear matter power spectrum from the linear one. This approach, although devised for $\Lambda$CDM, has been validated for dark energy models with similar matter power spectra, including varying $w(a)$ models, non-minimally coupled scalar fields~\citep{Dentler:2021zij}, massive neutrinos, and Vainshtein-screened modified gravity theories \citep{Mead:2016zqy,Mead:2020vgs}, and is valid for 'well-behaved' UDM fast transition models with a smooth, oscillation-free $\Lambda$CDM limit. In contrast, for models where the power spectrum exhibits large oscillatory behaviour, the HMCODE procedure breaks down. In such regimes, the evolution of overdensities is not hierarchical, and the linear density contrast briefly acquires large values at certain scales, making the notion of a non-linear threshold ill-defined. 
%As shown in Fig.~\ref{powerspectra_matter}, the amplitude of the matter power spectrum significantly deviates from $\Lambda$CDM even on linear scales, leading to low-likelihood values in the linear analysis.
However, we find that applying non-linear corrections to an oscillatory linear power spectrum further enhances these oscillations, often leading to a rejection of such models, reflected in low likelihood values. This means that applying the HMCODE in this regime does not bias the result, since it does not promote the acceptance of models with large oscillations. We can thus safely run the analyses in the defined range of parameters (cf. Table~\ref{tab: udm_priors}) without additional cuts.

\subsection{Analysis}

With the setup described above, we performed six analyses using the PolyChord algorithm: CMB, WL ‘linear’, and WL ‘full’, for both UDM and $\Lambda$CDM. The $\Lambda$CDM ones are control analyses, made in the same conditions as the UDM cases, to provide a reference result for comparison.

%{\color{red} The parameter space consists of the physical baryon matter density, the amplitude of the primordial power spectrum, the spectral index, the Hubble constant, the fluid density ($\Omega_{\rm cdm}$ in the case of $\Lambda$CDM or $\Omega_{\rm UDM}$ in the case of UDM), the two additional UDM parameters ($a_{\rm t}$ and $\beta$), and the nuisance parameters $A_{\rm Planck}$ for CMB data, the amplitude of the intrinsic alignment model $A_{\rm IA}$, and a baryon feedback parameter $c_{\rm min}$ for the WL 'full' analysis. Other cosmological and nuisance parameters were kept fixed.}

The constraints found for all the parameters in the various analyses %for the six analyses (CMB, WL ‘linear’, and WL ‘full’ for both UDM and $\Lambda$CDM) 
are given in Tables~\ref{tab: UDM_results} and~\ref{tab: LCDM_UDM results}. Constraints are also shown for a few derived parameters including: $\Omega_{\rm m}$, that for UDM is defined as the first term of the right-hand side of Eq.~\eqref{rhoudm}, %$\Omega_{\rm \Lambda}$ (for $\Lambda$CDM),
$\sigma_8$; $S_8 = \sigma_8/\Omega_m$, and the maximum of the speed of sound, in the case of UDM. The evidences for the two models computed for each dataset are also given in the tables.

\begin{table}
%\centering
\caption{Marginalized 2-$\sigma$ credible intervals found for each of the cosmological and nuisance parameters in the parameter space considered for each of the three analyses of the UDM model. Constraints on derived parameters are also shown, as well as the evidence (ln(Ev)) computed for the purpose of model comparison.}
%\vspace*{0.5cm}
\begin{tabular} {|l|c|c|c|}
\hline
 UDM &  Planck-2018 & KV-450-linear & KV-450-full\\
\hline
%{\boldmath${\rm log}(-a_t)$} & $3.5^{+1.5}_{-1.7}$ & $3.3^{+1.8}_{-2.0}$ & $2.19^{+0.44}_{-0.51}      $\\
{\boldmath$a_{\rm t}       $} & $0.0022^{+0.016}_{-0.0025} $ & $0.0071^{+0.12}_{-0.0077}$ & $0.0015^{+0.0066}_{-0.0041} $\\

{\boldmath${\rm log}(\beta)   $} & $8.0^{+3.8}_{-3.4}$ & $6.7^{+4.8}_{-4.7}$ & $8.2^{+3.7}_{-3.1}      $\\

{\boldmath$\Omega_{\Lambda{\rm UDM}}$} & $0.683^{+0.016}_{-0.016}$ & $0.73^{+0.18}_{-0.20}$ & $0.76^{+0.14}_{-0.17}      $\\

{\boldmath$\omega_b        $} & $0.02239^{+0.00029}_{-0.00028}$ & $0.0224^{+0.0035}_{-0.0034}$ & $0.0225^{+0.0033}_{-0.0033}$ \\

{\boldmath${\rm ln}(10^{10}A_s)   $} & $3.045^{+0.011}_{-0.011}$ & $2.9^{+1.7}_{-1.3}$ & $3.3^{+1.6}_{-1.5}         $\\

{\boldmath$n_s            $} & $0.9664^{+0.0085}_{-0.0080}$ & $1.01^{+0.27}_{-0.29}$ & $1.07^{+0.24}_{-0.28}      $\\

{\boldmath$h              $} & $0.673^{+0.012}_{-0.012}$ & $0.725^{+0.090}_{-0.084}$ & $0.736^{+0.090}_{-0.097}   $\\

\hline

{\boldmath$\Omega_{\rm m}        $} & $0.316^{+0.016}_{-0.016}$ & $0.27^{+0.20}_{-0.18}$ & $0.24^{+0.17}_{-0.14}      $\\

{\boldmath$\sigma_8         $} & $0.797^{+0.030}_{-0.13}$ & $0.58^{+0.38}_{-0.45}$ & $0.85^{+0.31}_{-0.30}      $\\

{\boldmath $S_8                        $} & $0.818^{+0.057}_{-0.085}$ & $0.53^{+0.29}_{-0.37}$ & $0.731^{+0.077}_{-0.073}   $\\

%{\boldmath$H_0             $} & $67.3^{+1.2}_{-1.2}$ & $73^{+9}_{-8}$ & $75.0^{+7.1}_{-9.9}        $\\

{\boldmath$ c^2_{\rm s,max}$   }            & $0.037^{+0.89}_{-0.042}$ & $0.023^{+0.83}_{-0.026}$ & $0.03^{+0.13}_{-0.11}            $\\

\hline

{\boldmath$A_{\rm Planck}       $} & $1.0006^{+0.0046}_{-0.0046}$ & & \\

{\boldmath$A_{\rm IA}           $} &  & $-0.1^{+4.5}_{-4.9}        $ & $0.8^{+1.4}_{-1.4}         $\\

{\boldmath$c_{\rm min}          $} &  &  & $2.51^{+0.62}_{-0.55}      $\\

%a\_transition\_udm        $ & $0.0022^{+0.016}_{-0.0025} $ & $0.0071^{+0.12}_{-0.0077}$ & $0.0078^{+0.011}_{-0.0066} $\\

\hline

{\boldmath${\rm ln(Ev)}        $} & $-529.27^{+0.35}_{-0.35}$ & $-11.49^{+0.12}_{-0.12}$ & $-97.39^{+0.16}_{-0.16}$\\     
\hline
\end{tabular}
\label{tab: UDM_results}
\end{table}

\begin{table}
%\centering
\caption{Marginalized 2-$\sigma$ credible intervals found for each of the cosmological and nuisance parameters in the parameter space considered for each of the three analyses of the $\Lambda$CDM model. Constraints on derived parameters are also shown, as well as the evidence (ln(Ev)) computed for the purpose of model comparison.}
%\vspace*{0.5cm}
\begin{tabular} {|l | c | c |c|}
\hline
 $\Lambda${\rm }CDM & Planck-2018 &  KV-450-linear & KV-450\\
\hline
{\boldmath$\Omega_{\rm cdm}      $} & $0.266^{+0.015}_{-0.014}   $ & $0.24^{+0.18}_{-0.14}$  & $0.21^{+0.17}_{-0.13}      $\\

{\boldmath$\omega_b        $} & $0.02235^{+0.00029}_{-0.00029}$ & $0.0226^{+0.0035}_{-0.0036}$ & $0.0223^{+0.0035}_{-0.0034}$\\

{\boldmath${\rm ln}(10^{10}A_s)   $} & $3.044^{+0.012}_{-0.011}   $ & $2.9^{+1.7}_{-1.3}$ & $3.2^{+1.5}_{-1.5}         $\\

{\boldmath$n_s            $} & $0.9647^{+0.0082}_{-0.0084}$ & $1.01^{+0.28}_{-0.29}      $ & $1.03^{+0.24}_{-0.25}      $\\

{\boldmath$h              $} & $0.673^{+0.011}_{-0.011}   $ & $0.740^{+0.084}_{-0.093}   $ & $0.738^{+0.078}_{-0.085}   $\\

%{\boldmath$Y_{\rm He}           $} & $0.24783^{+0.00012}_{-0.00012}$\\

\hline

%{\boldmath$\Omega_\Lambda   $} & $0.683^{+0.015}_{-0.016}   $ &  & $0.75^{+0.13}_{-0.17}      $\\

{\boldmath$\Omega_{\rm m}        $} & $0.317^{+0.016}_{-0.015}   $ & $0.25^{+0.17}_{-0.15}      $ & $0.25^{+0.17}_{-0.13}      $\\

{\boldmath$\sigma_8         $} & $0.8113^{+0.0097}_{-0.010} $ & $0.70^{+0.35}_{-0.30}      $ & $0.84^{+0.29}_{-0.27}      $\\

{\boldmath$S_8         $} & $0.834^{+0.030}_{-0.030} $ &   $0.62^{+0.26}_{-0.23}$    & $0.736^{+0.064}_{-0.073}      $\\

%{\boldmath$H_0           $} & $67.3^{+1.1}_{-1.1}        $ & $74^{+8}_{-9}              $ & $73.8^{+7.8}_{-8.5}        $\\

\hline

{\boldmath$A_{\rm Planck}       $} & $1.0006^{+0.0049}_{-0.0046}$ & &\\

{\boldmath$A_{\rm IA}           $} &  & $0.1^{+4.5}_{-5.0}         $ & $0.9^{+1.1}_{-1.2}         $\\

{\boldmath$c_{\rm min}          $} &  &  & $2.49^{+0.57}_{-0.47}      $\\

\hline

{\boldmath${\rm ln(Ev)}        $} & $-528.09^{+0.15}_{-0.15}$  & $-12.09^{+0.15}_{-0.15}   $  & $-96.74^{+0.07}_{-0.07}$ \\   

\hline
\end{tabular}
\label{tab: LCDM_UDM results}
\end{table}

The UDM WL analysis using only linear scales %was done to avoid using a non-linear prescription and get a preliminary probing of the UDM models. It is clear that it 
has less constraining power. In particular, it is not capable of providing a meaningful estimate of $S_8$ because the $(\sigma_8,\Omega_{\rm m})$ contour is wide, not showing the well-known ‘banana shape’. %We do not further discuss its results, and leave the corresponding contour plots for the Appendix~\ref{Appendix: UDM Fast}, for completeness. {\color{red} Should we include this?}

\begin{figure*}
	\centering
\includegraphics[width=16cm]{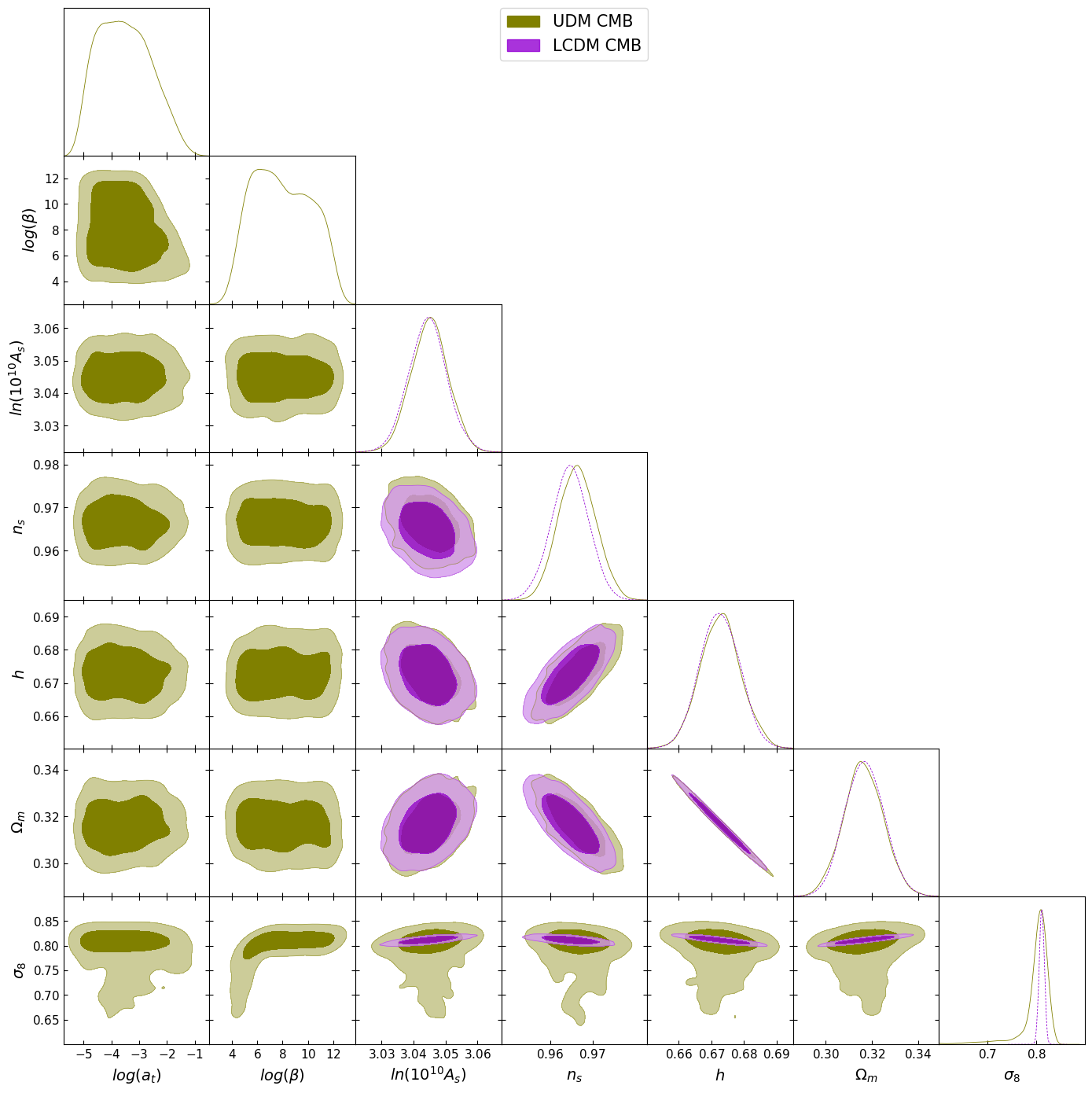}
	\caption {CMB analyses. Marginalized two-dimensional 1- and 2-$\sigma$ contours of the posterior and one-dimensional marginalized posterior for a subset of varied and derived parameters, for both $\Lambda$CDM (red) and UDM (green).}
\label{UDM_LCDM_CMB}
\end{figure*}

\begin{figure*}
	\centering
\includegraphics[width=16cm]{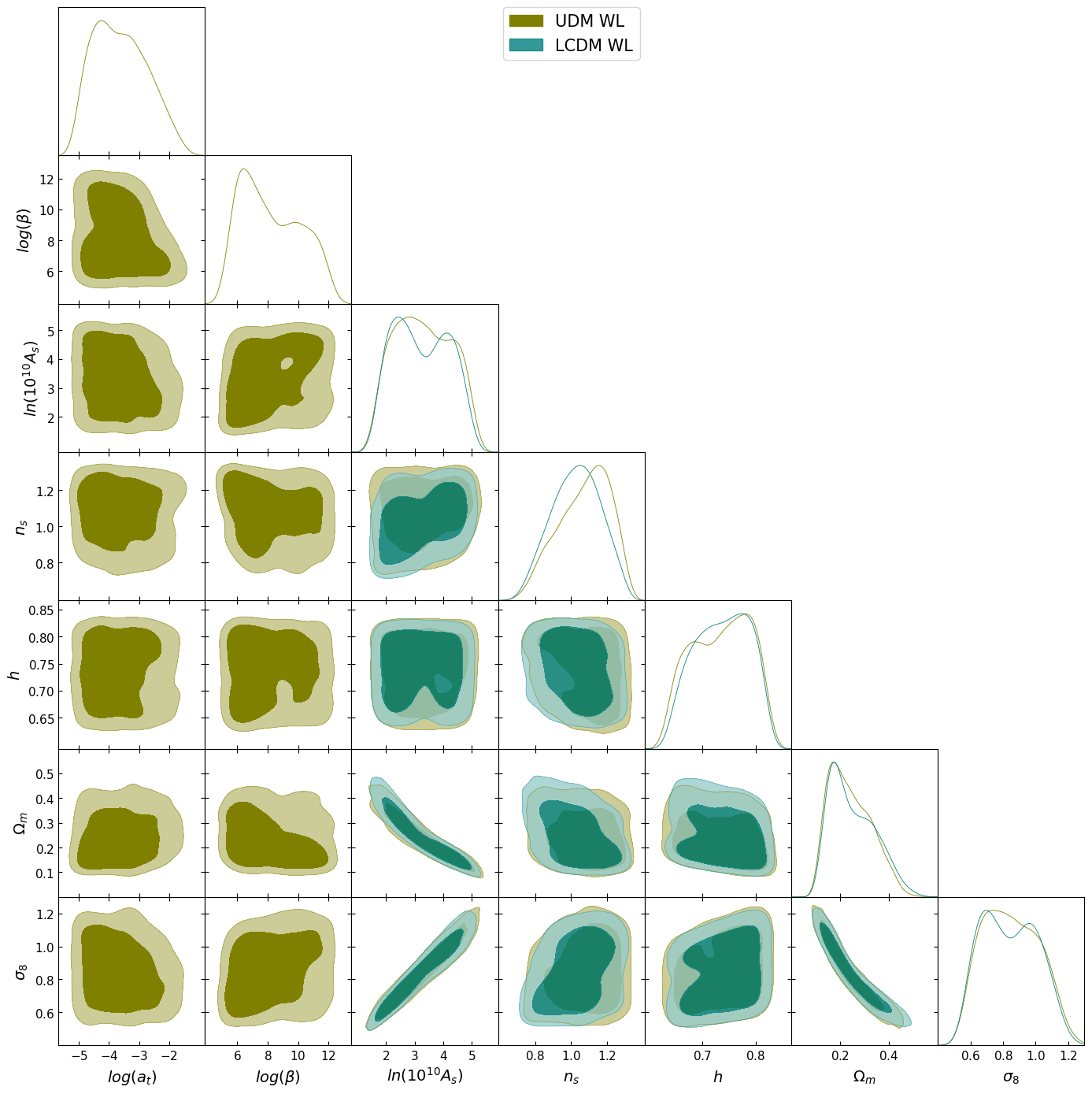}
	\caption {KiDS-full analyses. Marginalized two-dimensional 1- and 2-$\sigma$ contours of the posterior and one-dimensional marginalized posterior for a subset of varied and derived parameters, for both $\Lambda$CDM (red) and UDM (green).}
\label{UDM_LCDM_WL}
\end{figure*}

We focus thus on the results of the CMB and WL ‘full’ analyses. Figures~\ref{UDM_LCDM_CMB} and~\ref{UDM_LCDM_WL} show the marginalized  1- and 2-$\sigma$ credible contours resulting from the CMB and WL ‘full’ analyses, respectively. The marginalized two-dimensional countours are shown for all pairs among 8 parameters. The parameters displayed are the two UDM parameters $a_{\rm t}$, $\beta$, one density parameter, $\Omega_{\rm m}$ instead of $\Omega_{\Lambda\rm UDM}$ to allow direct comparison with $\Lambda$CDM, $n_{\rm s}$, $h$, the two amplitude parameters $A_{\rm s}$ and $\sigma_8$ and the combination $S_8$.

We observe that both datasets favour early transitions, finding a mean $a_{\rm t}$ of order $10^{-3}$. The $\beta$ one-dimensional distribution is skewed with a tail of significant values extending to its upper limit in both datasets. 

The CMB UDM analysis shows a correlation between $\beta$ and $\sigma_8$, with models with low $\beta$ values allowing for low $\sigma_8$ values. This is reflected in the $\sigma_8$ one-dimensional distribution, which shows a tail towards lower values. The CMB constraints for $\Lambda$CDM are similar to the UDM ones for parameters other than $\sigma_8$. The WL constraints on $a_{\rm t}$ and $\beta$ are not looser than the CMB ones. In contrast, the WL constraints on the standard parameters are an order of magnitude weaker than the CMB ones.

\begin{figure*}
	\centering
\includegraphics[width=16cm]{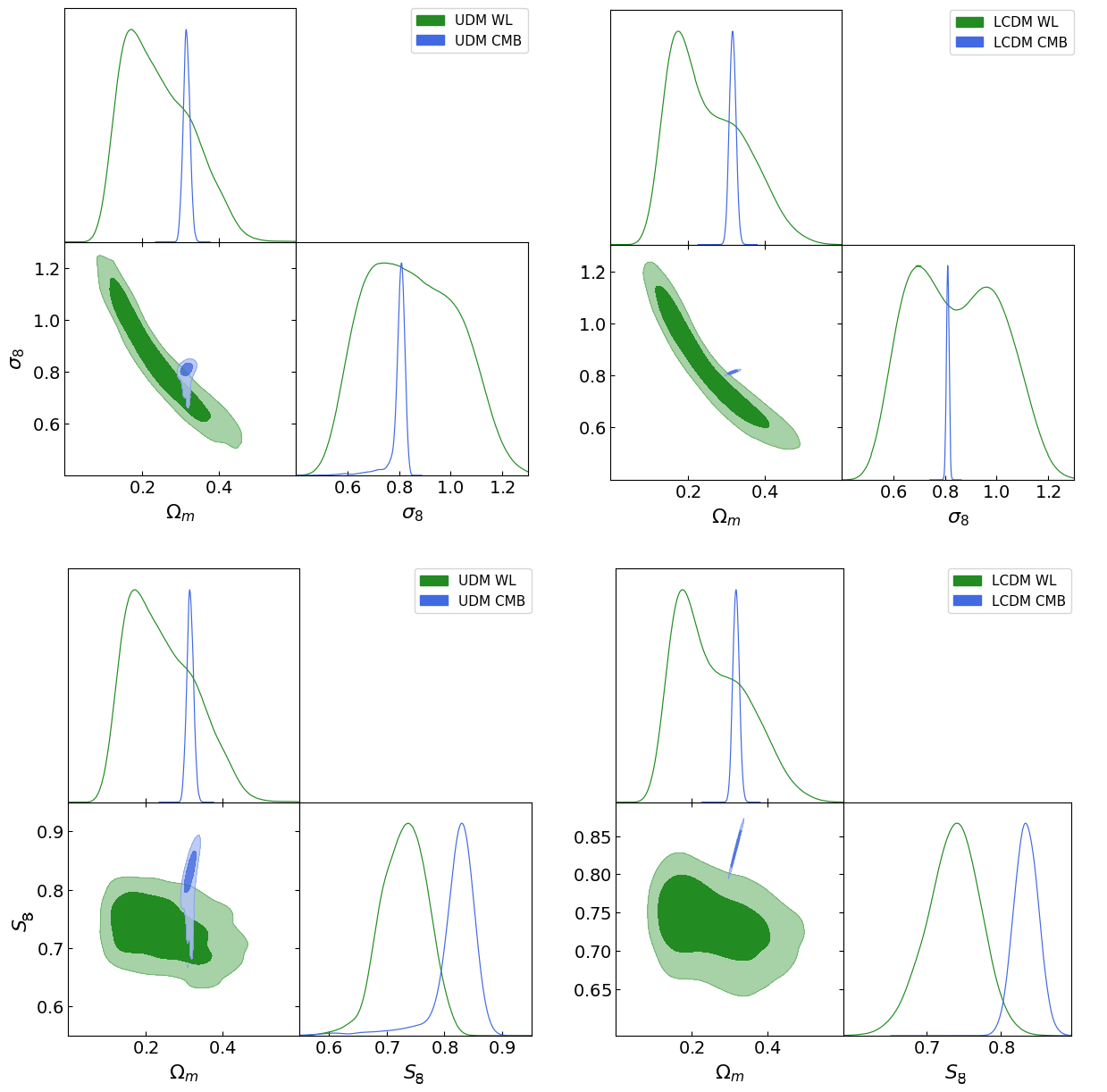}
	\caption {Marginalized one and two-dimensional posteriors constraints from the CMB (blue) and WL (green) analyses. Upper quadrants: ($\sigma_8$,$\Omega_{\rm m})$ plane. Lower  quadrants: ($S_8$,$\Omega_{\rm m})$ plane. Left quadrants: UDM. Right quadrants: $\Lambda$CDM.
    }
\label{UDM_Om_sigma8}
\end{figure*}

Figure~\ref{UDM_Om_sigma8} focuses on the $\sigma_8$ tension. It shows the marginalized  $(\Omega_{\rm m}, \sigma_8)$ and $(\Omega_{\rm m}, S_8)$ contours for each model, overlaying the results from the two probes for direct visualization of the tension.

The presence of the $\sigma_8$ tail in the UDM CMB results decreases the mean value of $S_8$ reducing its difference to the $S_8$ WL value by 10$\%$, when compared to the $\Lambda$CDM CMB-WL $S_8$ difference. On the other hand, the UDM combined CMB and WL 1-$\sigma$ constraints are 20$\%$ broader than in the $\Lambda$CDM case. The combination of the two factors leads to a decrease in the $S_8$ tension. We indeed find that the UDM model shows a $1.9\,\sigma$ $S_8$ tension between the datasets, while for $\Lambda$CDM we find a larger value of $2.6\,\sigma$, consistent with the value found in the literature for these particular datasets \citep{Hildebrandt:2018yau}. The $S_8$ tension is currently mostly resolved in the latest KiDS-Legacy analysis \citep{2025A&A...703A.158W}.

Looking at the evidences quoted in Tables~\ref{tab: UDM_results} and \ref{tab: LCDM_UDM results}, the corresponding Bayes factors of $\Lambda$CDM over UDM are 1.18 from CMB and 0.65 from WL 'full'. Following the Jeffreys' scale we conclude that the CMB dataset weakly favours $\Lambda$CDM over UDM, while the WL 'full' dataset is inconclusive \citep{Trotta:2008qt}.

\section{Discussion}

We found that the data favour an early-time transition occurring soon after recombination. 
This transition is associated with a large value of the rapidity parameter $\beta$, which appears to slightly improve the agreement between early- and late-time observations concerning the $S_8$ tension. 

Although the peak of the transition occurs after recombination, the Heaviside approximation may start to deviate from zero before recombination. Therefore, the second term in Eq.~\eqref{rhoudm}, which is negative before $a_{\rm t}$, may turn on before recombination, decreasing the amplitude of the UDM mean density. As a result, it can contribute at a level that, while small, is not negligible given the precision of Planck observations, thereby introducing slight deviations from the standard $\Lambda$CDM predictions. 

Our analysis expands previous studies of this UDM model based on background data \citep{Lazkoz:2016hmh, Leanizbarrutia:2017afj}, by implementing the model at the level of linear perturbations in the Boltzmann code \texttt{CLASS} and testing it against updated observational datasets. Compared with earlier analyses, our results show that the transition must have a significantly higher rapidity than previously inferred from background-only observations in order for the model to remain observationally viable, and that it must occur during the matter-dominated epoch. 

This had been previously suggested in \citep{Bruni:2012sn}, where the authors analysed the evolution of the density parameters $\Omega_{r}$ and $\Omega_{\rm m}$ in an Einstein-de Sitter Universe; $\Omega_{r}$, $\Omega_{\rm m}$, and $\Omega_{\Lambda}$ in the standard $\Lambda$CDM scenario; and the corresponding parameters in a UDM model. They showed that, if the UDM transition occurs too late, the matter-radiation equality epoch closely resembles that of a purely CDM-dominated model, occurring significantly earlier than in the standard $\Lambda$CDM case. Moreover, under such conditions, the effective cosmological constant in the UDM model would become dominant at a later epoch compared to the standard $\Lambda$CDM scenario.
This would have an impact on the matter power spectrum, since the slowing down of the perturbation growth due to the accelerated expansion would happen later than in $\Lambda$CDM, leading to too much power on small scales. This behaviour is also found in related studies where the authors analysed sharp transitions in the equation of state \citep{Jaber:2017bpx, Martins:2019ebg}. While these works constrain specific model classes, they confirm that such transitions are tightly constrained, and that both low- and high-redshift data push any viable transition to high redshifts, deep into the matter era.

We have also compared the UDM model with $\Lambda$CDM, both with and without the inclusion of the \textit{halofit} non-linear corrections. Within the region of parameter space where the deviations from $\Lambda$CDM are small and the oscillations in the matter power spectrum are negligible, the UDM model is able to form structure and produces results similar to those obtained in $\Lambda$CDM. These results show that the unified dark matter--dark energy paradigm still contains viable models, and encourage further investigation of this class of scenarios. Moreover, when tested against early-time data, the transition in the equation of state allows the model to accommodate a lower value of $\sigma_8$ for a fixed $\Omega_{\rm m}$, slightly improving the agreement with late-time observations compared with the corresponding $\Lambda$CDM results.

Although it is not possible to make a strong claim at this stage, since $\Lambda$CDM remains statistically favoured, our results motivate further studies of this type of UDM model. Recent works \citep [see e.g.][]{Camera:2017tws, Wang:2024rus, Yao:2024kex} have also shown promising indications that some UDM models could potentially alleviate the tension between CMB observations and the growth of large-scale structure, even though current datasets still provide overall support for $\Lambda$CDM. It is worth noting that 
%{\color{red} In addition, 
the UDM model with fast transition introduced in \citep{Bertacca:2010mt} was recently confronted with background data in \citep{Frion:2023xwq}, with the Bayesian analysis showing that the model did not significantly alleviate the Hubble tension and that the statistical comparison still favoured $\Lambda$CDM.

There are several directions along which the analysis could be extended. One important direction concerns the non-linear regime. In this work, we made use of linear-to-non-linear mappings that were built for $\Lambda$CDM, and we were careful to apply them only in the region of the parameter space where UDM models show small deviations from $\Lambda$CDM and negligible oscillations in the matter power spectrum. Nevertheless, a more consistent treatment would require a UDM-specific non-linear solution, for example through dedicated $N$-body simulations. This will be particularly important if this type of model is to be tested with high-precision data from the Euclid space mission.

Another direction would be to consider the backreaction effect. This effect is relevant for single-fluid models \citep{Avelino:2014nva}, and consists in removing the collapsed fraction from the background dynamics of the fluid, effectively recovering part of the behaviour of a two-fluid description.

Finally, there is also the possibility of considering other types of UDM models in which the oscillations are limited by construction. This can be done by 
considering a Lagrangian formulation for this model, for example following the prescription of \citep{Bertacca:2010mt}. This allows a more generalised fluid description with non-adiabatic perturbations. Alternatively, the non-adiabaticity can be introduced directly in the fluid without recurring to a non-adiabatic scalar field \citep{2013PhRvD..88b3505Y}. These approaches would introduce a new degree of freedom, since the pressure perturbation would then depend not only on the density perturbation but also on entropy perturbations, allowing the effective sound speed to be controlled independently and potentially reducing the oscillations.

Although these approaches would introduce additional degrees of freedom, which could penalise the model according to information criteria, they would allow UDM models to be explored in a wider context. This may help identify which features of this class of models are capable of addressing the tensions present in $\Lambda$CDM, and may ultimately provide a better understanding of the dynamics of the dark sector.

\section*{Acknowledgements}

This work was supported by Funda\c{c}\~ao para a Ci\^encia e a Tecnologia (FCT) through national funds under the research grant UID/04434/2025 (DOI 10.54499/UID/04434/2025).
The authors acknowledge the FCT project “BEYLA– BEYond LAmbda" with ref. number PTDC/FIS-AST/0054/2021 and DC was supported by IDPASC through the grant No. PD/BD/150489/2019.
%support from Funda\c{c}\~ao para a Ci\^encia e a Tecnologia (FCT) through research grant UID/FIS/04434/2013. IT is funded through the Investigador FCT research contract IF/01518/2014 that also partially supported DC. ARF gratefully acknowledges support from FCT through Fellowship SFRH/BPD/96981/2103, and DC thanks IA through Fellowship No. IA2018-33-BIM. 
We thank Fabien Koehlinger and Thejs Brinckmann for clarifications about the KiDS likelihood and MontePython, respectively. DC also thanks Vasco Ferreira %and Katrine Marques 
for very useful discussions.

%%%%%%%%%%%%%%%%%%%%%%%%%%%%%%%%%%%%%%%%%%%%%%%%%%



\bibliographystyle{aa}
\bibliography{CCUDMFASTSF} % if your bibtex file is called example.bib

@misc{Lesgourgues:2011re,
    author = "Lesgourgues, Julien",
    title = "{The Cosmic Linear Anisotropy Solving System (CLASS) I: Overview}",
    eprint = "1104.2932",
    archivePrefix = "arXiv",
    primaryClass = "astro-ph.IM",    
    month = "4",
    year = "2011"
}

@article{Wetterich:1987fm,
    author = "Wetterich, C.",
    title = "{Cosmology and the Fate of Dilatation Symmetry}",
    eprint = "1711.03844",
    archivePrefix = "arXiv",
    primaryClass = "hep-th",
    reportNumber = "PRINT-87-0756, DESY-87-123",
    doi = "10.1016/0550-3213(88)90193-9",
    journal = "Nucl. Phys. B",
    volume = "302",
    pages = "668--696",
    year = "1988"
}

@article{Ratra:1987rm,
    author = "Ratra, Bharat and Peebles, P. J. E.",
    title = "{Cosmological Consequences of a Rolling Homogeneous Scalar Field}",
    reportNumber = "PUPT-1072",
    doi = "10.1103/PhysRevD.37.3406",
    journal = "Phys. Rev. D",
    volume = "37",
    pages = "3406",
    year = "1988"
}

@article{Caldwell:1997ii,
    author = "Caldwell, R. R. and Dave, Rahul and Steinhardt, Paul J.",
    title = "{Cosmological imprint of an energy component with general equation of state}",
    eprint = "astro-ph/9708069",
    archivePrefix = "arXiv",
    doi = "10.1103/PhysRevLett.80.1582",
    journal = "Phys. Rev. Lett.",
    volume = "80",
    pages = "1582--1585",
    year = "1998"
}

@article{Zlatev:1998tr,
    author = "Zlatev, Ivaylo and Wang, Li-Min and Steinhardt, Paul J.",
    title = "{Quintessence, cosmic coincidence, and the cosmological constant}",
    eprint = "astro-ph/9807002",
    archivePrefix = "arXiv",
    doi = "10.1103/PhysRevLett.82.896",
    journal = "Phys. Rev. Lett.",
    volume = "82",
    pages = "896--899",
    year = "1999"
}

@article{Caldwell:2005tm,
    author = "Caldwell, R. R. and Linder, Eric V.",
    title = "{The Limits of quintessence}",
    eprint = "astro-ph/0505494",
    archivePrefix = "arXiv",
    doi = "10.1103/PhysRevLett.95.141301",
    journal = "Phys. Rev. Lett.",
    volume = "95",
    pages = "141301",
    year = "2005"
}

@article{Armendariz-Picon:2000nqq,
    author = "Armendariz-Picon, C. and Mukhanov, Viatcheslav F. and Steinhardt, Paul J.",
    title = "{A Dynamical solution to the problem of a small cosmological constant and late time cosmic acceleration}",
    eprint = "astro-ph/0004134",
    archivePrefix = "arXiv",
    doi = "10.1103/PhysRevLett.85.4438",
    journal = "Phys. Rev. Lett.",
    volume = "85",
    pages = "4438--4441",
    year = "2000"
}

@article{Armendariz-Picon:2000ulo,
    author = "Armendariz-Picon, C. and Mukhanov, Viatcheslav F. and Steinhardt, Paul J.",
    title = "{Essentials of k essence}",
    eprint = "astro-ph/0006373",
    archivePrefix = "arXiv",
    doi = "10.1103/PhysRevD.63.103510",
    journal = "Phys. Rev. D",
    volume = "63",
    pages = "103510",
    year = "2001"
}

@article{Wetterich:1994bg,
    author = "Wetterich, Christof",
    title = "{The Cosmon model for an asymptotically vanishing time dependent cosmological 'constant'}",
    eprint = "hep-th/9408025",
    archivePrefix = "arXiv",
    reportNumber = "HD-THEP-94-16",
    journal = "Astron. Astrophys.",
    volume = "301",
    pages = "321--328",
    year = "1995"
}

@article{Amendola:1999er,
    author = "Amendola, Luca",
    title = "{Coupled quintessence}",
    eprint = "astro-ph/9908023",
    archivePrefix = "arXiv",
    doi = "10.1103/PhysRevD.62.043511",
    journal = "Phys. Rev. D",
    volume = "62",
    pages = "043511",
    year = "2000"
}

@article{Capozziello:2003tk,
    author = "Capozziello, Salvatore and Carloni, Sante and Troisi, Antonio",
    title = "{Quintessence without scalar fields}",
    eprint = "astro-ph/0303041",
    archivePrefix = "arXiv",
    journal = "Recent Res. Dev. Astron. Astrophys.",
    volume = "1",
    pages = "625",
    year = "2003"
}

@article{Carroll:2003wy,
    author = "Carroll, Sean M. and Duvvuri, Vikram and Trodden, Mark and Turner, Michael S.",
    title = "{Is cosmic speed - up due to new gravitational physics?}",
    eprint = "astro-ph/0306438",
    archivePrefix = "arXiv",
    reportNumber = "FERMILAB-PUB-03-263-A, SU-GP-03-6-2",
    doi = "10.1103/PhysRevD.70.043528",
    journal = "Phys. Rev. D",
    volume = "70",
    pages = "043528",
    year = "2004"
}

@article{Nojiri:2010wj,
    author = "Nojiri, Shin'ichi and Odintsov, Sergei D.",
    title = "{Unified cosmic history in modified gravity: from F(R) theory to Lorentz non-invariant models}",
    eprint = "1011.0544",
    archivePrefix = "arXiv",
    primaryClass = "gr-qc",
    doi = "10.1016/j.physrep.2011.04.001",
    journal = "Phys. Rept.",
    volume = "505",
    pages = "59--144",
    year = "2011"
}

@article{Dvali:2000hr,
    author = "Dvali, G. R. and Gabadadze, Gregory and Porrati, Massimo",
    title = "{4-D gravity on a brane in 5-D Minkowski space}",
    eprint = "hep-th/0005016",
    archivePrefix = "arXiv",
    reportNumber = "NYU-TH-00-04-01",
    doi = "10.1016/S0370-2693(00)00669-9",
    journal = "Phys. Lett. B",
    volume = "485",
    pages = "208--214",
    year = "2000"
}

@article{Deffayet:2000uy,
    author = "Deffayet, Cedric",
    title = "{Cosmology on a brane in Minkowski bulk}",
    eprint = "hep-th/0010186",
    archivePrefix = "arXiv",
    reportNumber = "NYU-TH-00-10-07",
    doi = "10.1016/S0370-2693(01)00160-5",
    journal = "Phys. Lett. B",
    volume = "502",
    pages = "199--208",
    year = "2001"
}

@article{Tomita:2000jj,
    author = "Tomita, Kenji",
    title = "{A local void and the accelerating universe}",
    eprint = "astro-ph/0011484",
    archivePrefix = "arXiv",
    doi = "10.1046/j.1365-8711.2001.04597.x",
    journal = "Mon. Not. Roy. Astron. Soc.",
    volume = "326",
    pages = "287",
    year = "2001"
}

@article{Alnes:2005rw,
    author = "Alnes, Havard and Amarzguioui, Morad and Gron, Oyvind",
    title = "{An inhomogeneous alternative to dark energy?}",
    eprint = "astro-ph/0512006",
    archivePrefix = "arXiv",
    doi = "10.1103/PhysRevD.73.083519",
    journal = "Phys. Rev. D",
    volume = "73",
    pages = "083519",
    year = "2006"
}

@article{Kamenshchik:2001cp,
    author = "Kamenshchik, Alexander Yu. and Moschella, Ugo and Pasquier, Vincent",
    title = "{An Alternative to quintessence}",
    eprint = "gr-qc/0103004",
    archivePrefix = "arXiv",
    doi = "10.1016/S0370-2693(01)00571-8",
    journal = "Phys. Lett. B",
    volume = "511",
    pages = "265--268",
    year = "2001"
}

@article{Bento:2002ps,
    author = "Bento, M. C. and Bertolami, O. and Sen, A. A.",
    title = "{Generalized Chaplygin gas, accelerated expansion and dark energy matter unification}",
    eprint = "gr-qc/0202064",
    archivePrefix = "arXiv",
    doi = "10.1103/PhysRevD.66.043507",
    journal = "Phys. Rev. D",
    volume = "66",
    pages = "043507",
    year = "2002"
}

@article{Sandvik:2002jz,
    author = "Sandvik, Havard and Tegmark, Max and Zaldarriaga, Matias and Waga, Ioav",
    title = "{The end of unified dark matter?}",
    eprint = "astro-ph/0212114",
    archivePrefix = "arXiv",
    doi = "10.1103/PhysRevD.69.123524",
    journal = "Phys. Rev. D",
    volume = "69",
    pages = "123524",
    year = "2004"
}

@article{Amendola:2005rk,
    author = "Amendola, Luca and Waga, Ioav and Finelli, Fabio",
    title = "{Observational constraints on silent quartessence}",
    eprint = "astro-ph/0509099",
    archivePrefix = "arXiv",
    doi = "10.1088/1475-7516/2005/11/009",
    journal = "JCAP",
    volume = "11",
    pages = "009",
    year = "2005"
}

@article{Kumar:2014pia,
    author = "Kumar, Sumit and Sen, Anjan A",
    title = "{Clustering GCG: a viable option for unified dark matter-dark energy?}",
    eprint = "1405.5688",
    archivePrefix = "arXiv",
    primaryClass = "astro-ph.CO",
    doi = "10.1088/1475-7516/2014/10/036",
    journal = "JCAP",
    volume = "10",
    pages = "036",
    year = "2014"
}

@article{Avelino:2014nva,
    author = "Avelino, P. P. and Bolejko, K. and Lewis, G. F.",
    title = "{Nonlinear Chaplygin Gas Cosmologies}",
    eprint = "1403.1718",
    archivePrefix = "arXiv",
    primaryClass = "astro-ph.CO",
    doi = "10.1103/PhysRevD.89.103004",
    journal = "Phys. Rev. D",
    volume = "89",
    number = "10",
    pages = "103004",
    year = "2014"
}

@article{Scherrer:2004au,
    author = "Scherrer, Robert J.",
    title = "{Purely kinetic k-essence as unified dark matter}",
    eprint = "astro-ph/0402316",
    archivePrefix = "arXiv",
    doi = "10.1103/PhysRevLett.93.011301",
    journal = "Phys. Rev. Lett.",
    volume = "93",
    pages = "011301",
    year = "2004"
}

@article{Piattella:2009kt,
    author = "Piattella, Oliver F. and Bertacca, Daniele and Bruni, Marco and Pietrobon, Davide",
    title = "{Unified Dark Matter models with fast transition}",
    eprint = "0911.2664",
    archivePrefix = "arXiv",
    primaryClass = "astro-ph.CO",
    doi = "10.1088/1475-7516/2010/01/014",
    journal = "JCAP",
    volume = "01",
    pages = "014",
    year = "2010"
}

@article{Bertacca:2010mt,
    author = "Bertacca, Daniele and Bruni, Marco and Piattella, Oliver F. and Pietrobon, Davide",
    title = "{Unified Dark Matter scalar field models with fast transition}",
    eprint = "1011.6669",
    archivePrefix = "arXiv",
    primaryClass = "astro-ph.CO",
    doi = "10.1088/1475-7516/2011/02/018",
    journal = "JCAP",
    volume = "02",
    pages = "018",
    year = "2011"
}

@article{Bruni:2012sn,
    author = "Bruni, Marco and Lazkoz, Ruth and Rozas-Fernandez, Alberto",
    title = "{Phenomenological models for Unified Dark Matter with fast transition}",
    eprint = "1210.1880",
    archivePrefix = "arXiv",
    primaryClass = "astro-ph.CO",
    doi = "10.1093/mnras/stt395",
    journal = "Mon. Not. Roy. Astron. Soc.",
    volume = "431",
    pages = "2907--2916",
    year = "2013"
}

@book{Mukhanov:2005sc,
    author = "Mukhanov, V.",
    title = "{Physical Foundations of Cosmology}",
    doi = "10.1017/CBO9780511790553",
    isbn = "978-0-521-56398-7",
    publisher = "Cambridge University Press",
    address = "Oxford",
    year = "2005"
}

@article{Lazkoz:2016hmh,
    author = "Lazkoz, Ruth and Leanizbarrutia, Iker and Salzano, Vincenzo",
    title = "{Cosmological constraints on fast transition unified dark energy and dark matter models}",
    eprint = "1602.01331",
    archivePrefix = "arXiv",
    primaryClass = "astro-ph.CO",
    doi = "10.1103/PhysRevD.93.043537",
    journal = "Phys. Rev. D",
    volume = "93",
    number = "4",
    pages = "043537",
    year = "2016"
}

@article{WMAP:2012fli,
    author = "Bennett, C. L. and others",
    collaboration = "WMAP",
    title = "{Nine-Year Wilkinson Microwave Anisotropy Probe (WMAP) Observations: Final Maps and Results}",
    eprint = "1212.5225",
    archivePrefix = "arXiv",
    primaryClass = "astro-ph.CO",
    doi = "10.1088/0067-0049/208/2/20",
    journal = "Astrophys. J. Suppl.",
    volume = "208",
    pages = "20",
    year = "2013"
}

@article{Haslam:1982zz,
    author = "Haslam, C. G. T. and Salter, C. J. and Stoffel, H. and Wilson, W. E.",
    title = "{A 408 MHz all-sky continuum survey. II. The atlas of contour maps}",
    journal = "Astron. Astrophys. Suppl. Ser.",
    volume = "47",
    pages = "1--142",
    year = "1982"
}

@article{Hildebrandt:2018yau,
    author = "Hildebrandt, H. and others",
    title = "{KiDS+VIKING-450: Cosmic shear tomography with optical and infrared data}",
    eprint = "1812.06076",
    archivePrefix = "arXiv",
    primaryClass = "astro-ph.CO",
    doi = "10.1051/0004-6361/201834878",
    journal = "Astron. Astrophys.",
    volume = "633",
    pages = "A69",
    year = "2020"
}

@article{Brinckmann:2018cvx,
    author = "Brinckmann, Thejs and Lesgourgues, Julien",
    title = "{MontePython 3: boosted MCMC sampler and other features}",
    eprint = "1804.07261",
    archivePrefix = "arXiv",
    primaryClass = "astro-ph.CO",
    reportNumber = "TTK-18-15",
    doi = "10.1016/j.dark.2018.100260",
    journal = "Phys. Dark Univ.",
    volume = "24",
    pages = "100260",
    year = "2019"
}

@article{Audren:2012wb,
    author = "Audren, Benjamin and Lesgourgues, Julien and Benabed, Karim and Prunet, Simon",
    title = "{Conservative Constraints on Early Cosmology: an illustration of the Monte Python cosmological parameter inference code}",
    eprint = "1210.7183",
    archivePrefix = "arXiv",
    primaryClass = "astro-ph.CO",
    reportNumber = "CERN-PH-TH-2012-290, LAPTH-048-12",
    doi = "10.1088/1475-7516/2013/02/001",
    journal = "JCAP",
    volume = "02",
    pages = "001",
    year = "2013"
}

@article{Lewis:2002ah,
    author = "Lewis, Antony and Bridle, Sarah",
    title = "{Cosmological parameters from CMB and other data: A Monte Carlo approach}",
    eprint = "astro-ph/0205436",
    archivePrefix = "arXiv",
    doi = "10.1103/PhysRevD.66.103511",
    journal = "Phys. Rev. D",
    volume = "66",
    pages = "103511",
    year = "2002"
}

@article{Feroz:2007kg,
    author = "Feroz, Farhan and Hobson, M. P.",
    title = "{Multimodal nested sampling: an efficient and robust alternative to MCMC methods for astronomical data analysis}",
    eprint = "0704.3704",
    archivePrefix = "arXiv",
    primaryClass = "astro-ph",
    doi = "10.1111/j.1365-2966.2007.12353.x",
    journal = "Mon. Not. Roy. Astron. Soc.",
    volume = "384",
    pages = "449",
    year = "2008"
}

@article{Feroz:2008xx,
    author = "Feroz, F. and Hobson, M. P. and Bridges, M.",
    title = "{MultiNest: an efficient and robust Bayesian inference tool for cosmology and particle physics}",
    eprint = "0809.3437",
    archivePrefix = "arXiv",
    primaryClass = "astro-ph",
    doi = "10.1111/j.1365-2966.2009.14548.x",
    journal = "Mon. Not. Roy. Astron. Soc.",
    volume = "398",
    pages = "1601--1614",
    year = "2009"
}

@article{Feroz:2013hea,
    author = "Feroz, F. and Hobson, M. P. and Cameron, E. and Pettitt, A. N.",
    title = "{Importance Nested Sampling and the MultiNest Algorithm}",
    eprint = "1306.2144",
    archivePrefix = "arXiv",
    primaryClass = "astro-ph.IM",
    doi = "10.21105/astro.1306.2144",
    journal = "Open J. Astrophys.",
    volume = "2",
    number = "1",
    pages = "10",
    year = "2019"
}

@article{Handley:2015fda,
    author = "Handley, W. J. and Hobson, M. P. and Lasenby, A. N.",
    title = "{PolyChord: nested sampling for cosmology}",
    eprint = "1502.01856",
    archivePrefix = "arXiv",
    primaryClass = "astro-ph.CO",
    doi = "10.1093/mnrasl/slv047",
    journal = "Mon. Not. Roy. Astron. Soc.",
    volume = "450",
    number = "1",
    pages = "L61--L65",
    year = "2015"
}

@article{Foreman-Mackey:2012any,
    author = "Foreman-Mackey, Daniel and Hogg, David W. and Lang, Dustin and Goodman, Jonathan",
    title = "{emcee: The MCMC Hammer}",
    eprint = "1202.3665",
    archivePrefix = "arXiv",
    primaryClass = "astro-ph.IM",
    doi = "10.1086/670067",
    journal = "Publ. Astron. Soc. Pac.",
    volume = "125",
    pages = "306--312",
    year = "2013"
}

@article{Akeret:2012ky,
    author = "Akeret, Joel and Seehars, Sebastian and Amara, Adam and Refregier, Alexandre and Csillaghy, Andre",
    title = "{CosmoHammer: Cosmological parameter estimation with the MCMC Hammer}",
    eprint = "1212.1721",
    archivePrefix = "arXiv",
    primaryClass = "astro-ph.CO",
    doi = "10.1016/j.ascom.2013.06.003",
    journal = "Astronomy and Computing",
    volume = "2",
    pages = "27--39",
    year = "2013"
    
}

@article{Leanizbarrutia:2017afj,
    author = "Leanizbarrutia, Iker and Rozas-Fern\'andez, Alberto and Tereno, Ismael",
    title = "{Cosmological constraints on a unified dark matter-energy scalar field model with fast transition}",
    eprint = "1706.01706",
    archivePrefix = "arXiv",
    primaryClass = "astro-ph.CO",
    doi = "10.1103/PhysRevD.96.023503",
    journal = "Phys. Rev. D",
    volume = "96",
    number = "2",
    pages = "023503",
    year = "2017"
}

@article{Jaber:2017bpx,
    author = "Jaber, Mariana and de la Macorra, Axel",
    title = "{Probing a Steep EoS for Dark Energy with latest observations}",
    eprint = "1708.08529",
    archivePrefix = "arXiv",
    primaryClass = "astro-ph.CO",
    doi = "10.1016/j.astropartphys.2017.11.007",
    journal = "Astropart. Phys.",
    volume = "97",
    pages = "130--135",
    year = "2018"
}

@article{Martins:2019ebg,
    author = "Martins, C. J. A. P. and Prat Colomer, M.",
    title = "{Fine-structure constant constraints on late-time dark energy transitions}",
    eprint = "1903.04310",
    archivePrefix = "arXiv",
    primaryClass = "astro-ph.CO",
    doi = "10.1016/j.physletb.2019.02.039",
    journal = "Phys. Lett. B",
    volume = "791",
    pages = "230--235",
    year = "2019"
}

@article{Wands:2012vg,
    author = "Wands, David and De-Santiago, Josue and Wang, Yuting",
    title = "{Inhomogeneous vacuum energy}",
    eprint = "1203.6776",
    archivePrefix = "arXiv",
    primaryClass = "astro-ph.CO",
    doi = "10.1088/0264-9381/29/14/145017",
    journal = "Class. Quant. Grav.",
    volume = "29",
    pages = "145017",
    year = "2012"
}

@book{Bracewell:2003,
    author = "Bracewell, R.",
    title = "{The Fourier Transform and Its Applications}",
    isbn = "978-0070582859",
    publisher = "McGraw-Hill",
    year = "2003"
}

@article{Beca:2005gc,
    author = "Beca, L. M. G. and Avelino, P. P.",
    title = "{Dynamics of perfect fluid Unified Dark Energy models}",
    eprint = "astro-ph/0507075",
    archivePrefix = "arXiv",
    doi = "10.1111/j.1365-2966.2007.11496.x",
    journal = "Mon. Not. Roy. Astron. Soc.",
    volume = "376",
    pages = "1169--1172",
    year = "2007"
}

@article{Mead:2016zqy,
    author = "Mead, Alexander and Heymans, Catherine and Lombriser, Lucas and Peacock, John and Steele, Olivia and Winther, Hans",
    title = "{Accurate halo-model matter power spectra with dark energy, massive neutrinos and modified gravitational forces}",
    eprint = "1602.02154",
    archivePrefix = "arXiv",
    primaryClass = "astro-ph.CO",
    doi = "10.1093/mnras/stw681",
    journal = "Mon. Not. Roy. Astron. Soc.",
    volume = "459",
    number = "2",
    pages = "1468--1488",
    year = "2016"
}

@article{Dentler:2021zij,
    author = {Dentler, Mona and Marsh, David J. E. and Hlo\v{z}ek, Ren\'ee and Lagu\"e, Alex and Rogers, Keir K. and Grin, Daniel},
    title = "{Fuzzy dark matter and the Dark Energy Survey Year 1 data}",
    eprint = "2111.01199",
    archivePrefix = "arXiv",
    primaryClass = "astro-ph.CO",
    reportNumber = "KCL-PH-TH/2021-84",
    doi = "10.1093/mnras/stac1946",
    journal = "Mon. Not. Roy. Astron. Soc.",
    volume = "515",
    number = "4",
    pages = "5646--5664",
    year = "2022"
}

@article{Mead:2020vgs,
    author = {Mead, Alexander and Brieden, Samuel and Tr\"oster, Tilman and Heymans, Catherine},
    title = "{hmcode-2020: improved modelling of non-linear cosmological power spectra with baryonic feedback}",
    eprint = "2009.01858",
    archivePrefix = "arXiv",
    primaryClass = "astro-ph.CO",
    doi = "10.1093/mnras/stab082",
    journal = "Mon. Not. Roy. Astron. Soc.",
    volume = "502",
    number = "1",
    pages = "1401--1422",
    year = "2021"
}

@ARTICLE{2025A&A...703A.158W,
       author = {{Wright}, Angus H. and {St{\"o}lzner}, Benjamin and {Asgari}, Marika and {Bilicki}, Maciej and {Giblin}, Benjamin and {Heymans}, Catherine and {Hildebrandt}, Hendrik and {Hoekstra}, Henk and {Joachimi}, Benjamin and {Kuijken}, Konrad and {Li}, Shun-Sheng and {Reischke}, Robert and {von Wietersheim-Kramsta}, Maximilian and {Yoon}, Mijin and {Burger}, Pierre and {Chisari}, Nora Elisa and {de Jong}, Jelte and {Dvornik}, Andrej and {Georgiou}, Christos and {Harnois-D{\'e}raps}, Joachim and {Jalan}, Priyanka and {William}, Anjitha John and {Joudaki}, Shahab and {Lesci}, Giorgio Francesco and {Linke}, Laila and {Loureiro}, Arthur and {Mahony}, Constance and {Maturi}, Matteo and {Miller}, Lance and {Moscardini}, Lauro and {Napolitano}, Nicola R. and {Porth}, Lucas and {Radovich}, Mario and {Schneider}, Peter and {Tr{\"o}ster}, Tilman and {Valentijn}, Edwin and {Wittje}, Anna and {Yan}, Ziang and {Zhang}, Yun-Hao},
        title = "{KiDS-Legacy: Cosmological constraints from cosmic shear with the complete Kilo-Degree Survey}",
      journal = {\aap},
         year = 2025,
        month = nov,
       volume = {703},
          eid = {A158},
        pages = {A158},
          doi = {10.1051/0004-6361/202554908},
archivePrefix = {arXiv},
       eprint = {2503.19441},
 primaryClass = {astro-ph.CO},
       adsurl = {https://ui.adsabs.harvard.edu/abs/2025A&A...703A.158W}
      }

@book{DE,
    author = {P.J. Steinhardt} ,
    title = "{Critical Problems in Physics, edited by
V.L. Fitch and D.R. Marlow}",
    publisher = "Princeton University" ,
    year = "1997"
}

@article{Chiba:1999ka,
    author = "Chiba, Takeshi and Okabe, Takahiro and Yamaguchi, Masahide",
    title = "{Kinetically driven quintessence}",
    eprint = "astro-ph/9912463",
    archivePrefix = "arXiv",
    reportNumber = "UTAP-352",
    doi = "10.1103/PhysRevD.62.023511",
    journal = "Phys. Rev. D",
    volume = "62",
    pages = "023511",
    year = "2000"
}

@article{Planck:2018vyg,
    author = "Aghanim, N. and others",
    collaboration = "Planck",
    title = "{Planck 2018 results. VI. Cosmological parameters}",
    eprint = "1807.06209",
    archivePrefix = "arXiv",
    primaryClass = "astro-ph.CO",
    doi = "10.1051/0004-6361/201833910",
    journal = "Astron. Astrophys.",
    volume = "641",
    pages = "A6",
    year = "2020",
    note = "[Erratum: Astron.Astrophys. 652, C4 (2021)]"
}

@article{Trotta:2008qt,
    author = "Trotta, Roberto",
    title = "{Bayes in the sky: Bayesian inference and model selection in cosmology}",
    eprint = "0803.4089",
    archivePrefix = "arXiv",
    primaryClass = "astro-ph",
    doi = "10.1080/00107510802066753",
    journal = "Contemp. Phys.",
    volume = "49",
    pages = "71--104",
    year = "2008"
}

@article{Ma:1995ey,
    author = "Ma, Chung-Pei and Bertschinger, Edmund",
    title = "{Cosmological perturbation theory in the synchronous and conformal Newtonian gauges}",
    eprint = "astro-ph/9506072",
    archivePrefix = "arXiv",
    doi = "10.1086/176550",
    journal = "Astrophys. J.",
    volume = "455",
    pages = "7--25",
    year = "1995"
}

@article{Camera:2017tws,
    author = "Camera, Stefano and Martinelli, Matteo and Bertacca, Daniele",
    title = "{Does quartessence ease cosmic tensions?}",
    eprint = "1704.06277",
    archivePrefix = "arXiv",
    primaryClass = "astro-ph.CO",
    doi = "10.1016/j.dark.2018.11.008",
    journal = "Phys. Dark Univ.",
    volume = "23",
    pages = "100247",
    year = "2019"
}

@article{Wang:2024rus,
    author = "Wang, Junchao and Huang, Zhiqi and Yao, Yanhong and Liu, Jianqi and Huang, Lu and Su, Yan",
    title = "{A PAge-like Unified Dark Fluid model}",
    eprint = "2405.05798",
    archivePrefix = "arXiv",
    primaryClass = "astro-ph.CO",
    reportNumber = "SYSU-SPA-2024",
    doi = "10.1088/1475-7516/2024/09/053",
    journal = "JCAP",
    volume = "09",
    pages = "053",
    year = "2024"
}

@ARTICLE{Yao:2024kex,
       author = {{Yao}, Yan-Hong and {Liu}, Jian-Qi and {Huang}, Zhi-Qi and {Wang}, Jun-Chao and {Su}, Yan},
        title = "{New unified dark sector model and its implications on the {\ensuremath{\sigma}}8 and S8 tensions}",
      journal = {\prd},
         year = 2025,
        month = jun,
       volume = {111},
       number = {12},
          eid = {123508},
        pages = {123508},
          doi = {10.1103/p51k-7rw2},
archivePrefix = {arXiv},
       eprint = {2409.04678},
 primaryClass = {astro-ph.CO},
       adsurl = {https://ui.adsabs.harvard.edu/abs/2025PhRvD.111l3508Y}
}

@ARTICLE{Frion:2023xwq,
       author = {{Frion}, Emmanuel and {Camarena}, David and {Giani}, Leonardo and {Miranda}, Tays and {Bertacca}, Daniele and {Marra}, Valerio and {Piattella}, Oliver F.},
        title = "{Bayesian analysis of a Unified Dark Matter model with transition: can it alleviate the H0 tension?}",
      journal = {The Open Journal of Astrophysics},
         year = 2024,
        month = mar,
       volume = {7},
          eid = {17},
        pages = {17},
          doi = {10.21105/astro.2307.06320},
archivePrefix = {arXiv},
       eprint = {2307.06320},
 primaryClass = {astro-ph.CO},
       adsurl = {https://ui.adsabs.harvard.edu/abs/2024OJAp....7E..17F}
}

@ARTICLE{2025PDU....4901965D,
       author = {{Di Valentino}, Eleonora and {Said}, Jackson Levi and {Riess}, Adam and {Pollo}, Agnieszka and {Poulin}, Vivian and {G{\'o}mez-Valent}, Adri{\`a} and {Weltman}, Amanda and {Palmese}, Antonella and {Huang}, Caroline D. and {van de Bruck}, Carsten and {Saraf}, Chandra Shekhar and {Kuo}, Cheng-Yu and {Uhlemann}, Cora and {Grand{\'o}n}, Daniela and {Paz}, Dante and {Eckert}, Dominique and {Teixeira}, Elsa M. and {Saridakis}, Emmanuel N. and {Colg{\'a}in}, Eoin {\'O}. and {Beutler}, Florian and {Niedermann}, Florian and {Bajardi}, Francesco and {Barenboim}, Gabriela and {Gubitosi}, Giulia and {Musella}, Ilaria and {Banik}, Indranil and {Szapudi}, Istvan and {Singal}, Jack and {Cases}, Jaume Haro and {Chluba}, Jens and {Torrado}, Jes{\'u}s and {Mifsud}, Jurgen and {Jedamzik}, Karsten and {Said}, Khaled and {Dialektopoulos}, Konstantinos and {Herold}, Laura and {Perivolaropoulos}, Leandros and {Zu}, Lei and {Galbany}, Llu{\'\i}s and {Breuval}, Louise and {Visinelli}, Luca and {Escamilla}, Luis A. and {Anchordoqui}, Luis A. and {Sheikh-Jabbari}, M.~M. and {Lembo}, Margherita and {Dainotti}, Maria Giovanna and {Vincenzi}, Maria and {Asgari}, Marika and {Gerbino}, Martina and {Forconi}, Matteo and {Cantiello}, Michele and {Moresco}, Michele and {Benetti}, Micol and {Sch{\"o}neberg}, Nils and {Akarsu}, {\"O}zg{\"u}r and {Nunes}, Rafael C. and {Bernardo}, Reginald Christian and {Ch{\'a}vez}, Ricardo and {Anderson}, Richard I. and {Watkins}, Richard and {Capozziello}, Salvatore and {Li}, Siyang and {Vagnozzi}, Sunny and {Pan}, Supriya and {Treu}, Tommaso and {Irsic}, Vid and {Handley}, Will and {Giar{\`e}}, William and {Murakami}, Yukei and {Banihashemi}, Abdolali and {Poudou}, Ad{\`e}le and {Heavens}, Alan and {Kogut}, Alan and {Domi}, Alba and {Lenart}, Aleksander {\L}ukasz and {Melchiorri}, Alessandro and {Vadal{\`a}}, Alessandro and {Amon}, Alexandra and {Rivera}, Alexander Bonilla and {Reeves}, Alexander and {Zhuk}, Alexander and {Bonanno}, Alfio and {{\"O}vg{\"u}n}, Ali and {Pisani}, Alice and {Talebian}, Alireza and {Abebe}, Amare and {Aboubrahim}, Amin and {Gonz{\'a}lez Mor{\'a}n}, Ana Luisa and {Kov{\'a}cs}, Andr{\'a}s and {Lymperis}, Andreas and {Papatriantafyllou}, Andreas and {Liddle}, Andrew R. and {Paliathanasis}, Andronikos and {Borowiec}, Andrzej and {Yadav}, Anil Kumar and {Yadav}, Anita and {Sen}, Anjan Ananda and {William}, Anjitha John and {Davis}, Anne Christine and {Shajib}, Anowar J. and {Walters}, Anthony and {Lonappan}, Anto Idicherian and {Chudaykin}, Anton and {Capodagli}, Antonio and {da Silva}, Antonio and {De Felice}, Antonio and {Racioppi}, Antonio and {Oficial}, Araceli Soler and {Montiel}, Ariadna and {Favale}, Arianna and {Bernui}, Armando and {Velasco}, Arrianne Crystal and {Heinesen}, Asta and {Bakopoulos}, Athanasios and {Chatzistavrakidis}, Athanasios and {Khanpour}, Bahman and {Sathyaprakash}, Bangalore S. and {Zgirski}, Bartek and {L'Huillier}, Benjamin and {Famaey}, Benoit and {Jain}, Bhuvnesh and {Zhang}, Bing and {Karmakar}, Biswajit and {Dragovich}, Branko and {Thomas}, Brooks and {Correa}, Carlos and {Boiza}, Carlos G. and {Marques}, Catarina and {Escamilla-Rivera}, Celia and {Tzerefos}, Charalampos and {Zhang}, Chi and {De Leo}, Chiara and {Pfeifer}, Christian and {Lee}, Christine and {Venter}, Christo and {Gomes}, Cl{\'a}udio and {Roque De bom}, Clecio and {Moreno-Pulido}, Cristian and {Iosifidis}, Damianos and {Grin}, Dan and {Blixt}, Daniel and {Scolnic}, Dan and {Oriti}, Daniele and {Dobrycheva}, Daria and {Bettoni}, Dario and {Benisty}, David and {Fern{\'a}ndez-Arenas}, David and {Wiltshire}, David L. and {Sanchez Cid}, David and {Tamayo}, David and {Valls-Gabaud}, David and {Pedrotti}, Davide and {Wang}, Deng and {Staicova}, Denitsa and {Totolou}, Despoina and {Rubiera-Garcia}, Diego and {Milakovi{\'c}}, Dinko and {Pesce}, Dominic W. and {Sluse}, Dominique and {Borka}, Du{\v{s}}ko and {Yusofi}, Ebrahim and {Giusarma}, Elena and {Terlevich}, Elena and {Tomasetti}, Elena and {Vagenas}, Elias C. and {Fazzari}, Elisa and {Ferreira}, Elisa G.~M. and {Barakovic}, Elvis and {Dimastrogiovanni}, Emanuela and {Holm}, Emil Brinch and {Mottola}, Emil and {{\"O}z{\"u}lker}, Emre and {Specogna}, Enrico and {Brocato}, Enzo and {Jensko}, Erik and {Enriquez}, Erika Antonette and {Bhatia}, Esha and {Bresolin}, Fabio and {Avila}, Felipe and {Bouch{\`e}}, Filippo and {Bombacigno}, Flavio and {Anagnostopoulos}, Fotios K. and {Pace}, Francesco and {Sorrenti}, Francesco and {Lobo}, Francisco S.~N. and {Courbin}, Fr{\'e}d{\'e}ric and {Hansen}, Frode K. and {Sloan}, Greg and {Farrugia}, Gabriel and {Lynch}, Gabriel and {Garcia-Arroyo}, Gabriela and {Raimondo}, Gabriella and {Lambiase}, Gaetano and {Anand}, Gagandeep S. and {Poulot}, Gaspard and {Leon}, Genly and {Kouniatalis}, Gerasimos and {Nardini}, Germano and {Cs{\"o}rnyei}, G{\'e}za and {Galloni}, Giacomo},
        title = "{The CosmoVerse White Paper: Addressing observational tensions in cosmology with systematics and fundamental physics}",
      journal = {Physics of the Dark Universe},
         year = 2025,
        month = sep,
       volume = {49},
          eid = {101965},
        pages = {101965},
          doi = {10.1016/j.dark.2025.101965},
archivePrefix = {arXiv},
       eprint = {2504.01669},
 primaryClass = {astro-ph.CO},
       adsurl = {https://ui.adsabs.harvard.edu/abs/2025PDU....4901965D}
}

@ARTICLE{2013PhRvD..88b3505Y,
       author = {{Yang}, Weiqiang and {Xu}, Lixin},
        title = "{Unified dark fluid with fast transition: Including entropic perturbations}",
      journal = {\prd},
         year = 2013,
        month = jul,
       volume = {88},
       number = {2},
          eid = {023505},
        pages = {023505},
          doi = {10.1103/PhysRevD.88.023505},
archivePrefix = {arXiv},
       eprint = {1311.5644},
 primaryClass = {astro-ph.CO},
       adsurl = {https://ui.adsabs.harvard.edu/abs/2013PhRvD..88b3505Y}
}

%\begin{thebibliography}{99}
%\end{thebibliography}

%%%%%%%%%%%%%%%%%%%%%%%%%%%%%%%%%%%%%%%%%%%%%%%%%%

%%%%%%%%%%%%%%%%% APPENDICES %%%%%%%%%%%%%%%%%%%%%

%\appendix


%If you want to present additional material which would interrupt the flow of the main %paper, it can be placed in an Appendix which appears after the list of references.

%%%%%%%%%%%%%%%%%%%%%%%%%%%%%%%%%%%%%%%%%%%%%%%%%%

% Don't change these lines
\label{lastpage}
\end{document}